\documentclass[10pt,conference]{IEEEtran}
\IEEEoverridecommandlockouts

\usepackage[numbers,sort]{natbib}
\usepackage[colorlinks=true,citecolor=blue,linkcolor=blue,urlcolor=blue]{hyperref}
\usepackage{amsmath,amssymb,amsfonts}
\usepackage{algorithmic}
\usepackage{graphicx}
\usepackage{textcomp}
\usepackage{xcolor}
\usepackage{bm}
\usepackage{makecell}
\usepackage[linesnumbered,ruled,vlined]{algorithm2e}
\usepackage{caption}
\usepackage{color}
\usepackage{stfloats}
\usepackage{booktabs}
\usepackage{multirow}
\usepackage{verbatim}
\usepackage{bm}
\usepackage{url} 
\usepackage{framed} 
\usepackage{threeparttable}    
\usepackage{balance}  
\usepackage{color, colortbl}
\usepackage{enumitem}
	
\definecolor{Gray}{gray}{0.9}
\definecolor{LightCyan}{rgb}{0.88,1,1}

\usepackage{microtype}
\def\BibTeX{{\rm B\kern-.05em{\sc i\kern-.025em b}\kern-.08em
    T\kern-.1667em\lower.7ex\hbox{E}\kern-.125emX}}

\newcommand{\todo}[1]{\textcolor{black}{#1}}

\begin{document}

\title{Understanding the Complexity and Its Impact on Testing in ML-Enabled Systems \\ {\large A Case Sutdy on Rasa}}


\author{
	\IEEEauthorblockN{
		Junming Cao\IEEEauthorrefmark{1}, 
		Bihuan Chen\IEEEauthorrefmark{1}, 
		Longjie Hu\IEEEauthorrefmark{1}, 
		Jie Gao\IEEEauthorrefmark{2} 
	    Kaifeng Huang\IEEEauthorrefmark{1}
        and Xin Peng\IEEEauthorrefmark{1}}
	\IEEEauthorblockA{\IEEEauthorrefmark{1}Fudan University, Shanghai, China}
	\IEEEauthorblockA{\IEEEauthorrefmark{2}Singapore University of Technology and Design, Singapore}
} 

\maketitle

\begin{abstract}
Machine learning (ML) enabled systems are emerging with recent breakthroughs in ML. A model-centric view is widely taken by the literature to focus only on the analysis~of~ML~models. However, only a small body of work takes a system view that~looks at how ML components work with the system and how they affect software engineering for ML-enabled systems. In this paper, we adopt this system view, and conduct~a~case study on Rasa~3.0,~an industrial dialogue system that has been widely adopted by various companies around~the world.~Our~goal~is~to~characterize~the~complexity of such a large-scale ML-enabled system~and~to~understand the~impact of the complexity on testing. Our study reveals~practical implications for software engineering for ML-enabled systems.
\end{abstract}



\section{introduction}

The recent advances in machine learning (ML) have attracted an increasing interest in applying ML across a breadth~of~business domains, e.g., self-driving cars, virtual assistants, robotics, and health care. According to the Global AI Adoption~Index~by IBM~\cite{ibmreport}, 35\% of companies around the world have~deployed AI in their business, while 42\% of companies are exploring AI. Such a trend has caused the emergence of ML-enabled systems which are composed of ML and non-ML components.~ML~components are often important, but usually only a part of many components in ML-enabled systems~\cite{Christian2022}.

The previous research on software engineering~for machine learning often takes a model-centric view that~focuses~only~on the analysis of ML models~\cite{Christian2022, Katie2020}. For~example,~many~advances have been made for DL model testing~(e.g.,~\cite{Pei2017, Tian2018, Sun2018, Aggarwal2019, Kim2019, Feng2020, Zhang2020, Dola2021, ml_testing}),~verification~(e.g.,~\cite{Paulsen2020a, Toledo2021, Paulsen2020b, Baluta2021, Singh2019}) and debugging (e.g., \cite{Ma2018, Li2020, odena19a,Tao2020}). Only a small~body~of work takes a holistic system view, e.g., architectural design \cite{Yokoyama2019, Serban2022}, technical debt~\cite{hidden_technical_debt, tang2021empirical}, ML component entanglement \cite{Zhang2016, nushi2017human, fix_that_fails}, feature interaction \cite{Abdessalem2018, Abdessalem2020, feature_interaction}, and model interactions in Apollo~\cite{pengFirstLookIntegration2020}. However, the lack of system-level understanding of ML-enabled systems may hide problems in engineering ML-enabled systems and hinder practical solutions.

In this paper, we adopt this system view, and conduct~a~case study on Rasa 3.0 \cite{rasa} to characterize the complexity~of~such~a large-scale ML-enabled system as well as to understand~the~impact of the complexity on testing. Rasa is a task-oriented~industrial dialogue system that has been widely used by various companies around~the world. Therefore, we believe Rasa is a good representative of real-world ML-enabled systems. 

We first investigate~the~complexity of Rasa at three levels.~At the system level, we explore how ML components~are~adopted across the modules in Rasa. We find that there are~\todo{23}~ML~models in \todo{15} ML components across \todo{6} modules. At the interaction level,~we~analyze how ML components interact with other~components in Rasa. We find that there are \todo{43} interaction patterns and \todo{230} interaction instances across \todo{4} major categories and \todo{8} inner categories. At the component level, we investigate how the code of ML components is composed by what kinds~of~code. We find that \todo{57.1\%} of the code inside components are data processing code, and there are \todo{8} composition patterns between data processing code and model usage code.

We then explore the impact of the complexity on testing~from two perspectives. From the testing practice perspective,~we~analyze how is the characteristic of test cases, and how well~they cope with the complexity. We find that the test coverage~of~component interactions is low because of the complexity from huge configuration space and from hidden component interactions. From the mutation testing perspective, we study how~is~the~bug-finding capability of test cases and test data (i.e., the data for testing models), and how well they cope with the complexity. 
We find that there may be many potential bugs in data processing code that can only be detected by test cases, due to the complexity from data processing code.
The capability of test data to kill mutants is limited because of the complexity from huge configuration space.

Based on our case study, we highlight practical~implications to improve software engineering for ML-enabled~systems. For example, the configuration space of ML-enabled systems should be tested adequately, and configuration suggestions~should~be provided to developers.  A general taxonomy of data processing code should be constructed, and then the maintaining~and~testing tools for it can be developed. More integration-level test cases should be created to cover component interactions. Test cases and test data should be used in combination to detect both non-ML specific and ML-specific bugs.

In summary, this paper makes the following contributions.

\begin{itemize}
\item We conduct an in-depth case study on Rasa to characterize its complexity~and the~impact of its complexity on testing.
\item We highlight practical implications to improve software engineering for ML-enabled systems.
\end{itemize}


\section{Background and Study Design}
We present the architecture of a typical task-oriented~dialogue systems, an overview of Rasa, and our study design.

\subsection{Architecture of a Typical Task-Oriented Dialogue System}

A task-oriented dialogue system (TDS) aims to assist~users~in performing specific tasks, such as restaurant booking~and~flight booking~\cite{multiwoz}. 
A pipeline-based TDS consists of four parts,~i.e., natural language understanding (NLU), dialogue state tracking (DST), dialogue policy (Policy) and natural language generation (NLG)~\cite{zhang2020recent}. 
NLU parses a user utterance~into~a~structured~semantic representation, including intent and slot-values.~The~intent is a high-level classification of the user utterance, such as \texttt{Inform} and \texttt{Request}. 
Slot-values are task-specific entities that are mentioned in the user utterance, such as restaurant~price range and location preference. 
After tokenization and featurization of the user utterance, NLU applies classification~models~to recognize intent, and named entity extraction models to extract slot-values. 
DST takes the entire history of the conversation, including both user utterances with predicted intents~and~slot-values and system responses, to estimate the current dialogue state, which is usually formulated as a sequential prediction~task \cite{williams2016dialogstate}.
Dialogue state is typically the probability distribution of user intent and slot-values till the current timestamp.~Given~the estimated dialogue state, Policy generates the next system~action, such as \texttt{Query Database} and \texttt{Utter Question}. 
As Policy determines a series of actions sequentially, sequential models such as Recurrent Neural Network (RNN) are applied. 
For actions that require a response, NLG converts the action into a natural language utterance, which is often considered as a sequence generation task~\cite{wen-etal-2015-semantically}.

\subsection{An Overview of Rasa 3.0}\label{sec:rasa_overview}

\begin{figure*}[!t]
\centering
\includegraphics[scale=0.60]{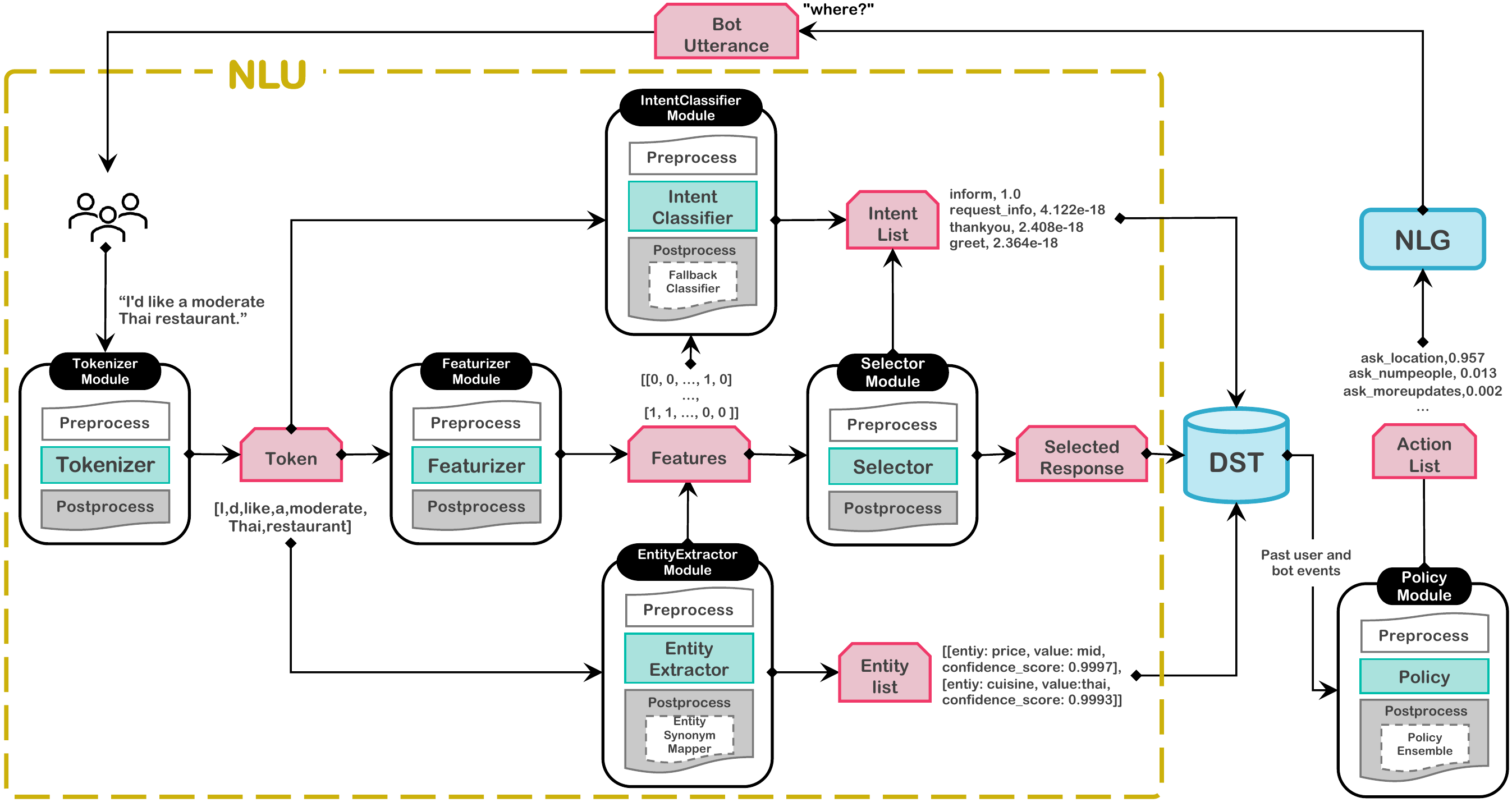}
\caption{The Modules and Workflow of Rasa}\label{overview_fig}
\end{figure*}

Rasa is a popular open-source ML-enabled TDS, which~is fully implemented with Python and used by many well-known companies in customer service for real users, including Adobe, Airbus, and N26 \cite{rasa}. 
An architecture overview~of~Rasa 3.0 is shown in Fig. \ref{overview_fig}. 
Each module consists of one single component or multiple semantically similar components. 
Apart from the modules in a typical TDS, Rasa proposes the Selector~module~to select candidate intents and responses for FAQ questions~\cite{chaudhuri-etal-2018-improving}.
We present some concepts in Rasa 3.0 to ease our presentation.

\textbf{Components in Rasa.} There are two types of components~in Rasa. We define \textit{ML components}~as~components that~are~implemented with ML models, and \textit{rule-based components} as components that are implemented with rule-based code logic.
General utils code in Rasa is not considered in this paper, such as command line and database access code.

\textbf{Configuration File and Component Graph in Rasa.}~As there are multiple available components~in~each~module,~developers need to choose components that are actually used~in the Rasa pipeline with a \textit{configuration file} to build a chatbot. 
Parameters of each component~are~specified~in~the~configuration file (e.g., ML model used by a component~and hyperparamers of a ML model).
Rasa applies Dask~\cite{dask} to compile~a~configuration file into a component graph. 
Each node in the component~graph denotes a component, and the edges connected with it denote upstream and downstream components with input and output data dependency. 
Execution of components obeys~the~topological order specified by edges.
These components interact with each other through fields in shared \texttt{Message} class instances. 
An upstream component stores outputs to \texttt{Message} instances, and a downstream component retrieves them for further processing. 

\textbf{ML Stages in Rasa.} 
Different from Apollo, which uses trained model files from external systems, and therefore~only contains the inference stage of ML models \cite{pengFirstLookIntegration2020}, the training, evaluation and inference stages of ML models are all~present in Rasa. 
Given a configuration file, Rasa separately compiles it to a training component graph and~an~inference~component graph. 
In training stage, the trainable~upstream~components are first trained, and then process the training data used by downstream components. 
In evaluation stage, only the performance metrics of \textit{IntentClassifier}, \textit{EntityExtractor} and \textit{Policy} are reported, as there is no ground truth for evaluation data in other modules.


\subsection{Study Design}

Our goal is to understand the complexity and its impact on testing in Rasa. To achieve this goal, we propose five RQs. 

\begin{itemize}[leftmargin=*]
    \item \textbf{RQ1 System Complexity Analysis}: how ML components~are adopted across the modules in Rasa?
    \item \textbf{RQ2 Interaction Complexity Analysis}: how ML components interact with other~components in Rasa?
    \item \textbf{RQ3 Component Complexity Analysis}: how the code of ML components is composed by what kinds~of~code?
    \item \textbf{RQ4 Testing Practice Analysis}: how is the characteristic of test cases, and how well~they cope with the complexity?
    \item \textbf{RQ5 Mutation Testing Analysis}: how~is~the~bug-finding capability of test cases and test data (i.e., the data for testing models), and how well they cope with the complexity?
\end{itemize}

\textbf{RQ1} aims to identify ML components in Rasa and broadly view them from the perspective of dependent libraries and ML models. 
\textbf{RQ2} aims to summarize a comprehensive taxonomy of component interaction patterns. 
\textbf{RQ3} aims to inspect the source code inside every component to characterize the statistics and composition patterns of different code types, including data processing code, model usage code, etc. 
Our findings from \textbf{RQ1}, \textbf{RQ2} and \textbf{RQ3} could reveal how the complexity originates and manifests in real world large-scale ML-enabled systems, which provide both practitioners and researchers with insights to overcome the complexities involved in implementing, maintaining, debugging and testing such complex systems. 

\textbf{RQ4} aims to quantitatively assess Rasa's test cases from~code coverage, test case statistics (i.e., granularity levels, oracle~types, and ML stages), and component interaction coverage perspectives. 
\textbf{RQ5} aims to generate mutants (i.e., artificial bugs) and check whether these mutants can be killed (i.e., detected) by test cases. 
Further, for the survived mutants, we train~Rasa~with 3 default configuration files on \textit{MutiWoz} \cite{multiwoz}, a widely used multi-domain TDS dataset. 
We calculate the statistical significance between the performance metrics from mutated Rasa code with metrics from pipelines trained with clean code.
Our findings from \textbf{RQ4} and \textbf{RQ5} evaluate the testing practice in Rasa, and shed light on automated test generation, bug localization and bug repairing techniques for complex ML-enabled systems.

\begin{table*}[!h]
	\caption{System Complexity Analysis of Rasa}
	\vspace{-10pt}
	\begin{center}
        \scalebox{0.97}{
	\begin{tabular}{llllllllll}
	\toprule
	\textbf{Module}
	& \textbf{Component}
	& \textbf{Direct Lib.}
	& \textbf{Indirect Lib.}
	& \textbf{Model Type}
	& \textbf{No. Model}
    & \textbf{Trainable}
    & \textbf{Rasa Imp.}
    & \textbf{LoC}\\
	\midrule
	\multirow{4}*{Tokenizer}
	&\cellcolor{Gray}JiebaTokenizer & \cellcolor{Gray}Jieba &\cellcolor{Gray}N/A &\cellcolor{Gray}HMM & \cellcolor{Gray}1 &\cellcolor{Gray}False  &\cellcolor{Gray}False   &\cellcolor{Gray}85  \\
	&\cellcolor{Gray}SpacyTokenizer &\cellcolor{Gray}Spacy &\cellcolor{Gray}Thinc &\cellcolor{Gray}MLP &\cellcolor{Gray}1 &\cellcolor{Gray}False  &\cellcolor{Gray}False &\cellcolor{Gray}39  \\
	&MitieTokenizer &Mitie &N/A &N/A &N/A &False  &False &43  \\
    &WhitespaceTokenizer &N/A &N/A &N/A &N/A &False &True &52    \\
	\midrule
	\multirow{7}*{Featurizer}
	&\cellcolor{Gray}ConveRTFeaturizer &\cellcolor{Gray}TensorFlow &\cellcolor{Gray}N/A &\cellcolor{Gray}Transformer &\cellcolor{Gray}1 &\cellcolor{Gray}False &\cellcolor{Gray}False &\cellcolor{Gray}269 \\
	&\cellcolor{Gray}LanguageModelFeaturizer &\cellcolor{Gray}Transformers &\cellcolor{Gray}TensorFlow &\cellcolor{Gray}Transformer &\cellcolor{Gray}6 &\cellcolor{Gray}False &\cellcolor{Gray}False &\cellcolor{Gray}378  \\
	&\cellcolor{Gray}MitieFeaturizer &\cellcolor{Gray}Mitie &\cellcolor{Gray}Dlib &\cellcolor{Gray}CCA &\cellcolor{Gray}1 &\cellcolor{Gray}False &\cellcolor{Gray}False &\cellcolor{Gray}98   \\
    &\cellcolor{Gray}SpacyFeaturizer &\cellcolor{Gray}Spacy &\cellcolor{Gray}Thinc &\cellcolor{Gray}CNN &\cellcolor{Gray}2 &\cellcolor{Gray}False &\cellcolor{Gray}False &\cellcolor{Gray}66   \\
    &CountVectorsFeaturizer &Scikit-learn &N/A &N/A &N/A &True &False &520   \\
    &LexicalSyntacticFeaturizer &N/A &N/A &N/A &N/A &True  &True &319  \\
    &RegexFeaturizer &N/A &N/A &N/A &N/A &True &True &151  \\
    \midrule
    \multirow{5}*{IntentClassifier}
    &\cellcolor{Gray}DIETClassifier &\cellcolor{Gray}TensorFlow &\cellcolor{Gray}N/A &\cellcolor{Gray}Transformer &\cellcolor{Gray}1 &\cellcolor{Gray}True &\cellcolor{Gray}True &\cellcolor{Gray}1217 \\
    &\cellcolor{Gray}MitieIntentClassifier &\cellcolor{Gray}Mitie &\cellcolor{Gray}Dlib &\cellcolor{Gray}SVM &\cellcolor{Gray}1 &\cellcolor{Gray}True &\cellcolor{Gray}False &\cellcolor{Gray}89 \\
    &\cellcolor{Gray}SklearnIntentClassifier &\cellcolor{Gray}Scikit-learn &\cellcolor{Gray}N/A &\cellcolor{Gray}SVM &\cellcolor{Gray}1 &\cellcolor{Gray}True &\cellcolor{Gray}False &\cellcolor{Gray}173 \\
    &FallbackClassifier  &N/A &N/A &N/A &N/A &False &True&91   \\
    &KeywordIntentClassifier  &N/A &N/A &N/A &N/A &True &True&132  \\
    \midrule
    \multirow{7}*{EntityExtractor}
    &\cellcolor{Gray}DIETClassifier &\cellcolor{Gray}TensorFlow &\cellcolor{Gray}N/A &\cellcolor{Gray}Transformer &\cellcolor{Gray}1 &\cellcolor{Gray}True &\cellcolor{Gray}True &\cellcolor{Gray}1217 \\
    &\cellcolor{Gray}CRFEntityExtractor &\cellcolor{Gray}Scikit-learn &\cellcolor{Gray}N/A &\cellcolor{Gray}CRF &\cellcolor{Gray}1 &\cellcolor{Gray}True &\cellcolor{Gray}False &\cellcolor{Gray}438 \\
    &\cellcolor{Gray}MitieEntityExtractor &\cellcolor{Gray}Mitie &\cellcolor{Gray}Dlib &\cellcolor{Gray}SVM &\cellcolor{Gray}1 &\cellcolor{Gray}True &\cellcolor{Gray}False &\cellcolor{Gray}164 \\
    &\cellcolor{Gray}SpacyEntityExtractor  &\cellcolor{Gray}Spacy &\cellcolor{Gray}Thinc &\cellcolor{Gray}MLP &\cellcolor{Gray}1 &\cellcolor{Gray}False  &\cellcolor{Gray}False &\cellcolor{Gray}52  \\
    &DucklingEntityExtractor  &N/A &N/A &N/A &N/A &False &False &134   \\
    &RegexEntityExtractor &N/A &N/A &N/A &N/A &True  &True &124  \\
    &EntitySynonymMapper &N/A &N/A &N/A &N/A &True  &True &102  \\
    \midrule
    Selector
    &\cellcolor{Gray}ResponseSelector &\cellcolor{Gray}TensorFlow &\cellcolor{Gray}N/A &\cellcolor{Gray}Transformer &\cellcolor{Gray}2 &\cellcolor{Gray}True &\cellcolor{Gray}True &\cellcolor{Gray}560  \\
    \midrule
    \multirow{6}*{Policy}
    &\cellcolor{Gray}TEDPolicy &\cellcolor{Gray}TensorFlow &\cellcolor{Gray}N/A &\cellcolor{Gray}Transformer+CRF &\cellcolor{Gray}1 &\cellcolor{Gray}True &\cellcolor{Gray}True &\cellcolor{Gray}1262 \\
    &\cellcolor{Gray}UnexpecTEDIntentPolicy &\cellcolor{Gray}TensorFlow &\cellcolor{Gray}N/A &\cellcolor{Gray}Transformer+CRF &\cellcolor{Gray}1 &\cellcolor{Gray}True &\cellcolor{Gray}True &\cellcolor{Gray}458 \\
    &MemoizationPolicy &N/A &N/A &N/A &N/A &True &True &207   \\
    &AugmentedMemoizationPolicy &N/A &N/A &N/A &N/A &True &True &65   \\
    &RulePolicy &N/A &N/A &N/A &N/A &True &True &818   \\
    &PolicyEnsemble &N/A &N/A &N/A &N/A &False &True &150   \\
	\bottomrule
	\end{tabular}		
	\label{system_complexity}}
	\end{center}
\end{table*}


\section{RQ1: System Complexity Analysis}

\subsection{Methodology}

To answer \textbf{RQ1}, we identified ML and rule-based components in Rasa and characterized them through a detailed examination of Rasa's source code and documentation. We excluded DST and NLG as they are fully implemented with rule-based code logic in Rasa without ML components. 

All the modules we identified are listed in Table \ref{system_complexity}, except for a special module, \textit{Shared}, as it contains general data processing code and  ML model definition code (e.g., Transformer), while does not contain any independent components. We will include it in the last three RQs. Specifically, for each component, we recursively tracked methods within it to manually extract the model or rule definition code. We examined implementation details of APIs in ML libraries by reading the documentations and source code of external libraries, including ML model type and number of candidate models. 

In particular, we analyzed whether ML components are implemented by using external direct libraries or indirect libraries, whether the components can be trained (notice that not only some of ML components can be trained, but also some rule-based components can be trained as long as they update internal parameters when processing training data),  whether Rasa implements components with its own code and provides built-in model and rule definition code, and the lines of code (LoC) of each component excluding blank lines, code comments and import statements. 






\subsection{Results}

The results are summarized in Table \ref{system_complexity}.
Components shown in gray color are ML components, and others are rule-based components.
There are 6 modules in total, including 15 ML and 14 rule-based components. These components contain 23 ML models and are implemented with 7 directly dependent external ML libraries and 3 indirectly dependent external ML libraries. In particular, all ML components in \textit{Tokenizer} and \textit{Featurizer} are not trainable because pre-trained language models are applied. All components in \textit{Policy} are implemented in Rasa's own code, because there are no ready-to-use Policy models provided by existing libraries. There are a total of 5348 LoC in ML components and 2980 LoC in rule-based components. In addition, the general module \textit{Shared} contains 5375 LoC, which is not listed in the table.

Notably, we find that classical machine learning models (e.g., Support Vector Machine and Conditional Random Field) together with deep learning models (e.g., Convolutional Neural Networks and Transformer) play an important role in Rasa. This is different from the previous study \cite{pengFirstLookIntegration2020} on Apollo, which is focused on deep learning models.

Next, we introduce components used in each module.

\textbf{Tokenizer.} Tokenizer splits the user utterance into tokens with component specific split symbols (e.g., whitespace and punctuation). (1) \textit{SpacyTokenizer} provides the richest token information, including splitting tokens with rules, lemmatizing tokens with a look-up table, and performing part-of-speech~tagging with a multi-layer perceptron (MLP). 
(2) \textit{JiebaTokenizer} is the only component that tokenizes non-English sentences using Hidden Markov Model (HMM) \cite{eddy1996hidden}. (3) \textit{MitieTokenizer} and \textit{WhitespaceTokenizer} toeknize text with predefined rules.

\textbf{Featurizer.} As shown in Fig. \ref{overview_fig}, Featurizer converts tokens into features for downstream module inference. (1) \textit{ConveRTFeaturizer} loads TFHub's \cite{TensorHub} pre-trained ConveRT (Conversational Representations from Transformers) TensorFlow model  to featurize tokens \cite{henderson2019convert}. (2) \textit{LanguageModelFeaturizer} loads pre-trained language models from Hugging Face Transformers \cite{transformers}, including BERT \cite{devlin2018bert}, GPT \cite{hu2020gpt}, XLNet \cite{yang2019xlnet}, Roberta \cite{liu2019roberta}, XLM \cite{xlm} and GPT2 \cite{gpt2}.
(3) \textit{MitieFeaturizer} combines Canonical Correlation Analysis (CCA) feature and word~morphology features together. (4) \textit{SpacyFeaturizer} applies HashEmbedCNN or Roberta to convert tokens to features, depending on the pre-trained Spacy pipeline specified in the configuration file.
(5) \textit{CountVectorsFeaturizer}, \textit{LexicalSyntacticFeaturizer} and \textit{RegexFeaturizer} create sparse features with n-grams, sliding window and regex patterns,  respectively. 


\textbf{IntentClassifier.} IntentClassifier generates a predicted intent list ordered by confidence scores based on tokens and features from upstream modules.
(1) \textit{DIETClassifier} implements Dual Intent and Entity Transformer (DIET) to perform intent~classification and entity recognition simultaneously, and is therefore included in both \textbf{IntentClassifier} and \textbf{EntityExtractor} modules. 
(2) \textit{MitieIntentClassifier} and \textit{SklearnIntentClassifier} apply a multi-class Support Vector Machine (SVM) \cite{Shmilovici2005} with a sparse linear kernel using Scikit-learn and Mitie, respectively.
(3) \textit{KeywordIntentClassifier} classifies  user intent with keywords extracted from training data.
(4) \textit{FallbackClassifier} is a post-processing component to check the results of other components in \textit{DIETClassifier}. It identifies a user utterance with the intent \texttt{nlu\_fallback} if the confidence scores are not greater than  \texttt{threshold}, or the score difference of the two highest ranked intents is less than the \texttt{ambiguity\_threshold}.

\textbf{EntityExtractor.} EntityExtractor extract entities such as the restaurant's location and price. (1) \textit{DIETClassifier} also serves as an EntityExtractor. 
(2) \textit{CRFEntityExtractor}, \textit{MitieEntityExtractor} and \textit{SpacyEntityExtractor} utilize a conditional random fields (CRF) model, a multi-class linear SVM, and a MLP to predict entities, respectively. 
(3) \textit{DucklingEntityExtractor} and \textit{RegexEntityExtractor} extract entities using a duckling server \cite{duckling} and regex patterns.
(4) \textit{EntitySynonymMapper} is a post-processing component to convert synonymous entity values~into a same value. As Fig. \ref{overview_fig} shows, the value of ``price" entity, ``moderate", is coverted to ``mid" by \textit{EntitySynonymMapper}.

\textbf{Selector.} \textit{ResponseSelector} aims to directly select the response from a set of candidate responses, which is also known as response selection task in the literature \cite{chaudhuri-etal-2018-improving}. It embeds user inputs and candidate responses in the same vector space, using the same neural network architecture as \textit{DIETClassifier}.

\textbf{Policy.} Policy decides the action the system takes on each conversation based on dialogue states.
(1) \textit{TEDPolicy} proposes a Transformer Embedding Dialogue (TED) model to embed dialogue states and system actions into a single~semantic~vector space, and select the action with the max similarity score~with the current dialogue states \cite{TED}.
(2) \textit{MemoizationPolicy}, \textit{AugmentedMemoizationPolicy} and \textit{RulePolicy} match the current conversation history with examples in the training data and predefined rules to predict system actions.
(3) \textit{UnexpecTEDIntentPolicy} decides on the possibility of the intent predicted by IntentClassifier given current dialogue states, which follows the same model architecture as \textit{TEDPolicy}.
(4) \textit{PolicyEnsemble} is a post-processing component to select the proper system action from output actions of different policies. 

\subsection{Implications}

The system complexity of Rasa poses challenges for developers using Rasa (i.e., application developers) and developers creating Rasa (i.e., system developers).

\textbf{Complexity from ML supply chain.} Rasa depends~on~10 external ML libraries directly or indirectly. Less than~100~(0.03\%) projects out of 355392 projects using TensorFlow on GitHub depend on 10 more DL libraries \cite{supply_chain}. It could be inferred that relying on 10 more ML libraries is also less common.
For application developers, it is difficult to understand the implementation details of components that rely on external ML libraries, not to mention selecting proper components and parameters. For example, due to the lack of documentations of \textit{MitieFeaturizer} in Rasa, application developers need to inspect Mitie's source code to learn that it implements CCA using Dlib APIs. 
For system developers, vulnerabilities \cite{npm_technical_lag} and dependency bugs \cite{dependency_bug} may arise because of outdated or incompatible library versions. 
Therefore, future work should provide supports for the management of components and corresponding dependent ML libraries for ML-enabled systems, similar to traditional software component analysis \cite{Foo2019TheDO}.

\textbf{Complexity from configurations.} It could be extremely complex to configure Rasa with 29 components and hundreds of parameters, making it easy to misconfigure and thus affect functionality and performance. This kind of misconfiguration is similar to what happens in traditional configurable software systems \cite{configurable_system}. Additionally, finding optimal configurations for application developers' specific TDS scenarios is difficult,~also known as configuration debt \cite{hidden_technical_debt}. Although AutoML~has~been extensively studied to select appropriate ML models~and~parameters for specific tasks, they all focus on selecting a single ML model without considering the combination of multiple ML models and rules  \cite{XinHe2021AutoMLAS}. Another challenge~is~to~detect~potential bugs by testing a huge set of configuration settings. Existing studies on ML model testing mainly focus on testing a single ML model with predefined hyperparameters \cite{ml_testing}.
 

\section{RQ2: Interaction Complexity Analysis}
\subsection{Methodology}

The workflow in Fig.~\ref{overview_fig} only shows the general~flow~among different modules. The details of interactions among components are still uncovered. We consider the interaction among two or more components, with at least one ML component.
An interaction pattern contains a module placeholder, which could be instantiated with components in the module~to~generate~interaction instances.
For example, pattern (PolicyEnsemble, [Policy]) could be instantiated as (PolicyEnsemble, TEDPolicy) or (PolicyEnsemble, (TEDPolicy, RulePolicy)). To answer \textbf{RQ2}, we conducted a qualitative and quantitative analysis of the component interaction patterns and instances of components.

\textbf{Step 1: Extract interaction patterns.}
The interaction can be divided into two categories: inter-module interaction and intra-module interaction.
(1) Inter-module interaction: the interaction between two adjacent modules (e.g., Featurizer with Tokenizer) was considered. We analyzed the usages~of~\texttt{Message} in the component code because components use the \texttt{Message} class to transfer data. Specifically, we extracted all interaction patterns of its upstream and downstream component. We also considered the interaction between post-processing component (i.e., \textit{FallbackClassifier}, \textit{EntitySynonymMapper} and \textit{PolicyEnsemble}) and other components in their residing modules as inter-module interaction.
(2) Intra-module interaction: we identify the interaction pattern for components in each module.

\textbf{Step 2: Generate interaction instances.}
For each inter-module interaction pattern, we instantiated the module placeholder with every component in the module. 
For every intra-module interaction, we extracted the Cartesian product~of~all components in each module as interaction instances.
We~then filtered out component instances that do not contain ML~components, or do not meet the constraints specified~in~Rasa~documentation. 
For example, \textit{CRFEntityExtractor} could not use features of \textit{SparseFeaturizer} other than \textit{RegexFeaturizer}.

\textbf{Step 3: Summarize the interaction taxonomy.} For generated component patterns and instances, we analyzed their semantics and summarized a component interaction taxonomy.

\subsection{Results}
\begin{figure}[!t]
    \centering
    \includegraphics[scale=0.38]{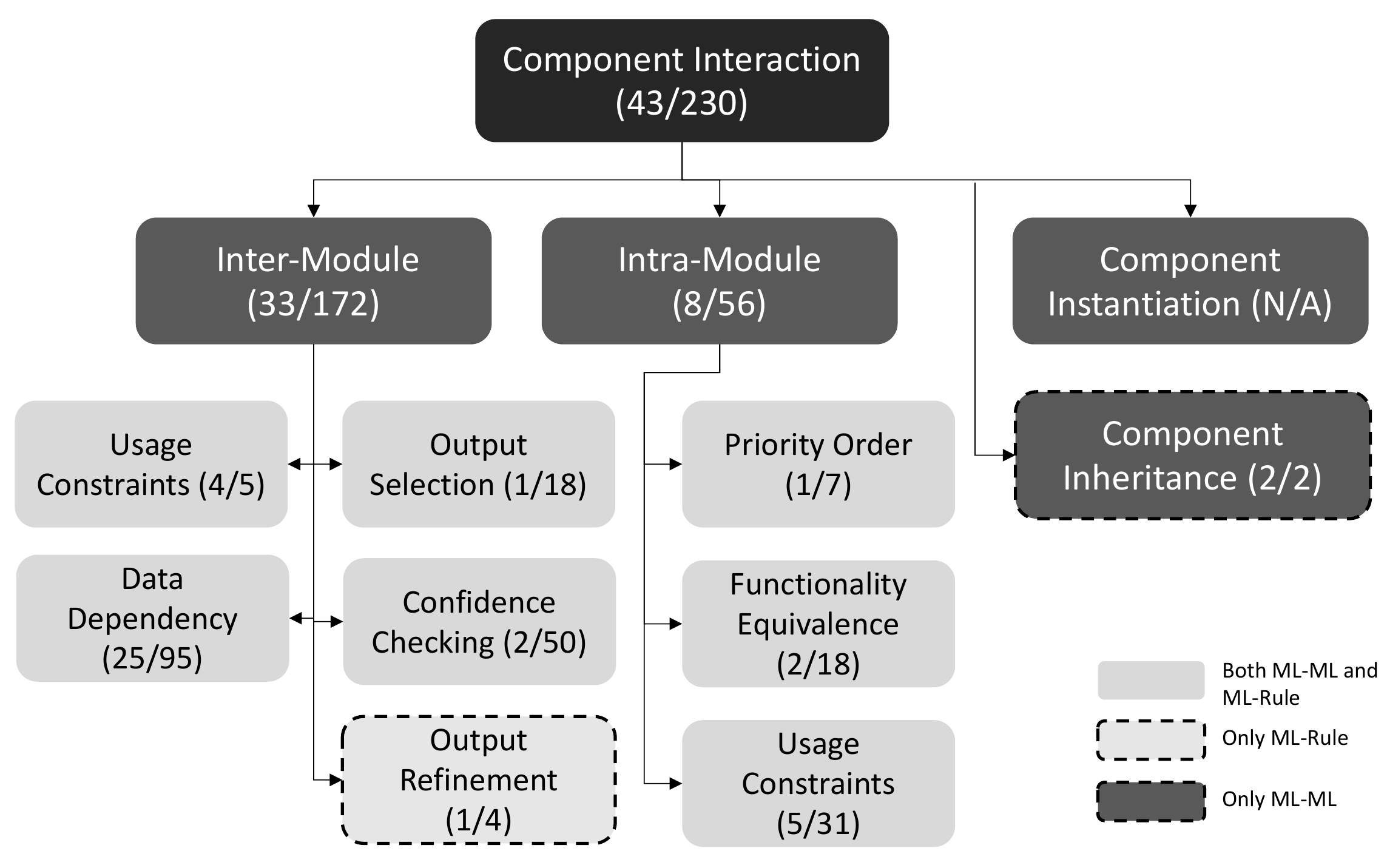}
    \caption{Taxonomy of Component Interactions}
    \label{component_interaction_fig}
\end{figure}

The component interaction taxonomy is shown in Fig. \ref{component_interaction_fig}. It is divided into 4 high-level categories (i.e. \textit{Inter-Module}, \textit{Intra-Module}, \textit{Component Instantiation} and \textit{Component Inheritance}) and 8 inner categories. 
Note that only \textit{Inter-Module} interactions contain components with direct data dependency, while other categories contain components with indirect interactions (e.g., two featurizers are used together). The number of interaction patterns and interaction instances in each category~is~listed~as \textit{pattern\_count}/\textit{instance\_count} in Fig. \ref{component_interaction_fig}. There are a total of 43 interaction patterns and 230 interaction instances.
Almost all categories include both ML to ML components and ML to rule-based components interactions. On the contrary, previous work on Apollo \cite{pengFirstLookIntegration2020} also presented 4 of the 8 inner categories, but did not provide a taxonomy and quantitative analysis. 


\textbf{Inter-Module}. Components from multiple modules interact through data transfers. In particular, \textit{Output Selection} means that the downstream component selects the proper output from multiple upstream outputs based on configurable criteria, e.g., \textit{PolicyEnsemble} with policies. 
\textit{Output Refinement} denotes that the downstream component complements the imperfect outputs of upstream components with rules, e.g., 
\textit{EntitySynonymMapper} with entity extractors. \textit{Confidence Checking} means that the downstream component checks reliability of the output from upstream components using ML models (e.g., \textit{UnexpecTEDIntentPolicy} with intent classifiers) or rules (e.g., \textit{FallbackClassifier} with IntentClassifiers). If the outputs are marked as not reliable, fallback behaviors such as the \texttt{fall\_back} system action are triggered. 
\textit{Usage Constraints} defines components that should or should not be used together under certain circumstances. For example, \textit{SpacyTokenizer} is required by \textit{CountVectorsFeaturizer} when applying \texttt{use\_lemma} option and \textit{LexicalSyntacticFeaturizer} when applying \texttt{pos\_tag} option. \textit{Data Dependency} includes the rest of inter-module interaction patterns that do not fall into any of the above categories, which are relatively ``trivial" interactions with no specific semantics.

\textbf{Intra-Module}. The interaction mode of components within a module differs from \textbf{Inter-Module}. These components interact indirectly when used together. 
\textit{Priority Order} means that the outputs of components within a module are selected according to priority order, e.g., the priority order of policies.
\textit{Usage Constraints} is similar to \textit{Usage Constraints} in the inter-module category. For example, only one component in any of \textit{Tokenizer}, \textit{IntentClassifier} and \textit{EntityExtractor} should be used in each configuration file, otherwise outputs of additional components will be overwritten.  \textit{Functionality Equivalence} includes all intra-module interaction patterns that do not belong to any of the above categories, which are relatively ``trivial" interactions involving components used together with no specific semantics.

\textbf{Component Instantiation}. Rasa supports creating multiple instances of a component within a configuration setting. For example, multiple \textit{CountVectorFeaturizers} instances with different ngram settings, and multiple \textit{LanguageModelFeaturizer} instances with different language models could be used together. We did not count this category of interaction patterns and instances, since developers could specify infinite instances of a component within a configuration setting.

\textbf{Component Inheritance}. The class inheritance mechanism allows ML models to be shared among components. For example, ML model definition class  in \textit{UnexpecTEDIntentPolicy} is a subclass of the ML model definition class in \textit{TEDPolicy}.

\subsection{Implications}


\textbf{Lack of specifications for interactions.} The outputs of ML components for specific inputs are not guaranteed due to the stochastic nature of ML models  \cite{feature_interaction}. Thus, it is more difficult to formulate interaction semantics in ML-enabled systems than in traditional systems. When testing samples are predicated wrongly, it is challenging to localize the exact faulty component. 
Moreover, even if the faulty component has been fixed and performance of it has been improved, the overall performance of the entire system may degrade \cite{fix_that_fails}. Therefore, training and evaluation should be extended from component-level to system-level to consider interactions among components. 
In summary, we need to pay more attention to addressing the challenges caused by lack of specifications in bug localization and repairing for ML-enabled systems.

\textbf{Hidden interactions}. It is non-trivial to identify all interactions even for system developers of Rasa. 
For example,~the \textit{Data Dependency} interaction between \textit{RegexFeaturizer} and \textit{CRFEntityExtractor} is not marked in documentations and can only be identified from source code.
Application developers may misuse components and get confused with the poor~performance of the system without understanding the hidden~interactions, especially for interaction categories like \textit{Usage Constraints}, \textit{Output Selection} and \textit{Priority Order}.
Techniques like~data flow analysis can be explored to automatically reveal component interactions in ML-enabled systems \cite{Sattler2017LiftingID}.

Furthermore, our results on component interaction complexity could be helpful to guide developers to build~a~better~ML-enabled system.
For example, developers can follow interaction patterns \textit{Output Selection} and \textit{Output Refinement} to improve the outputs of components at system level, as well as utilizing  \textit{Confidence Checking} to detect cases that ML models~can~not handle, and then triggering fallback rules, which is very important in safety-critical systems like self-driving systems~\cite{pengFirstLookIntegration2020}.




\section{RQ3: Component Complexity Analysis}

\subsection{Methodology}

To answer \textbf{RQ3}, we classified categories of code snippets in each component and explored their composition patterns. 

\textbf{Step 1: Label code snippets.} We segmented each source code file into code snippets according to semantic meaning, and then classified them into 6 categories: (1) model definition, the definition code of ML models; (2) rule definition, the definition code of rules in rule-based components; (3) model usage, the usage code of ML models; (4) rule usage, the usage code of rules; (5) data pre-processing, the input data processing code before model or rule usages; (6) data post-processing, the output data processing code after model or rule usages. Two of the authors labeled code snippets independently, and the third author was involved to resolve disagreements. The Cohen's Kappa coefficient of the two authors reached 0.830.

\textbf{Step 2: Summarize composition patterns of code snippets.} 
Based on labeled code snippets, we summarized the composition patterns of data processing code, and model or rule usage code in each component.  

\subsection{Results}
\begin{table}[!t]
	\caption{LoC of Different Code Categories}
	\vspace{-10pt}
    \footnotesize
	\begin{center}
    \tabcolsep=1.0mm
	\begin{tabular}{ccccccc}
        \hline
        \multirow{2}{*}{\textbf{Module}} & \multicolumn{2}{c}{\textbf{Data}} & \multicolumn{2}{c}{\textbf{Model}} & \multicolumn{2}{c}{\textbf{Rule}} \\ \cline{2-7} 
                                              & Pre. & Post.                 & Usage & Definition                  & Usage & Definition \\ \hline
        \multicolumn{1}{c|}{Tokenizer}        &8      & \multicolumn{1}{c|}{80} &27       & \multicolumn{1}{c|}{0} &25       &25      \\
        \multicolumn{1}{c|}{Featurizer}       &390      & \multicolumn{1}{c|}{323} &92       & \multicolumn{1}{c|}{0} &162       &119      \\
        \multicolumn{1}{c|}{IntentClassifier} &441      & \multicolumn{1}{c|}{131} &113       & \multicolumn{1}{c|}{298} &3       &69      \\
        \multicolumn{1}{c|}{EntityExtractor}  &746      & \multicolumn{1}{c|}{311} &120       & \multicolumn{1}{c|}{298} &24       &30      \\
        \multicolumn{1}{c|}{Selector}         &48      & \multicolumn{1}{c|}{55} &9       & \multicolumn{1}{c|}{16} &0       &0      \\
        \multicolumn{1}{c|}{Policy}           &1332     & \multicolumn{1}{c|}{540} &64       & \multicolumn{1}{c|}{543} &167       &283      \\
        \multicolumn{1}{c|}{Shared}           &996      & \multicolumn{1}{c|}{314} &112       & \multicolumn{1}{c|}{1673} &0       &43      \\
        \multicolumn{1}{c|}{Total}            &3961      & \multicolumn{1}{c|}{1754} &537       & \multicolumn{1}{c|}{2828} &381       &569      \\ \hline
        \end{tabular}
	\label{code_type_stat}
	\end{center}
\end{table}

The statistics of different code categories are shown in Table \ref{code_type_stat}. We only considered the LoC of labeled code snippets, while ignoring general utils code such as class initialization. 
Data processing code contributes a total of 5715 (57.1\%) LoC, while model usage\&definition code and rule usage\&definition code contribute 3365 (33.5\%) and 950 (9.4\%) LoC, respectively. 
1673 (59.2\%) of the 2828 LoC of model definition code is in  \textit{Shared} module, which shows that the reuse of model definition code between different components is quite common. There is no model definition code in \textit{Tokenizer} and \textit{Featurizer}, because ML components are all built on top of external ML libraries.


We classified data pre-processing and data post-processing categories into more specific types, due to the dominant proportion of data processing code in Rasa. 
Specifically, \textit{Validation} code intends to validate the input or output data of components. \textit{Format Transformation} code transforms data format, such as constructing vectors from Python arrays and reshaping vectors. 
\textit{Component Input/Output Filter} code filters data that does not meet the specified criteria, such as the absence of certain attributes. 
\textit{Data Scale/Padding/Encoding/Decoding} code changes the value of data, while \textit{Data Split/Shuffle/Balance/Batch/Rank} code changes the organization of data for better training and inference of components.
We provide the complete codebook and statistics of data processing types at our website \cite{website}.

\begin{figure*}[ht]
    \centering
    \includegraphics[scale=0.53]{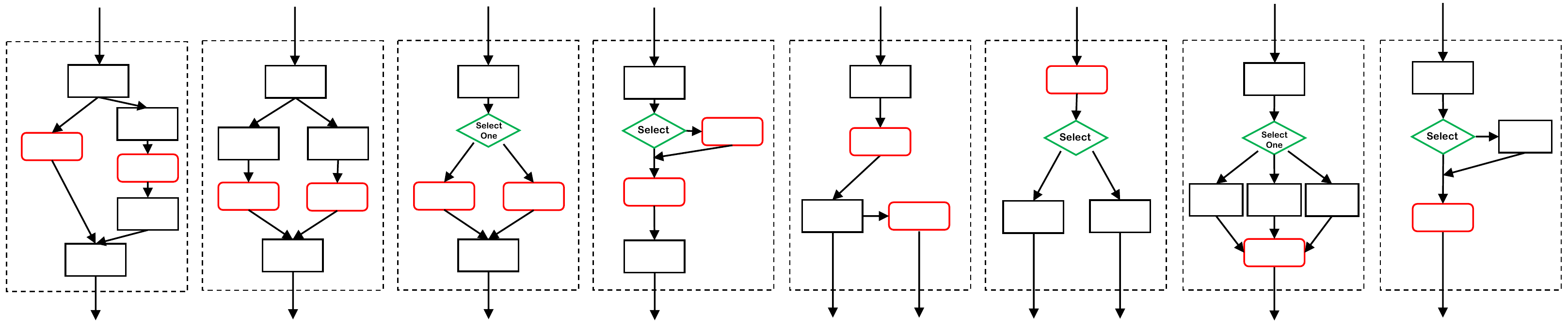}
    \vspace{-10pt}
    \caption{Non-Sequential Code Composition Patterns within Components}
    \label{code_composition}
\end{figure*}

Moreover, we find that composition patterns of code snippets include sequential code composition pattern and various non-sequential composition patterns.
In a typical sequential composition pattern, data is first pre-processed, and then processed by model or rule usage code, and finally post-processed.
The non-sequential code composition patterns are summarized in Fig. \ref{code_composition}. The black box is a data processing code snippet, the red box is a model or rule usage code snippet, and the green diamond means to select one or multiple downstream code snippets.
The first 5 patterns consist of multiple model or rule usages in one component. 
The last 3 patterns consist of a single model or rule usage with multiple possible data processing snippets, decided by configurations or input data.


\subsection{Implications}

\textbf{Data processing.} Data processing code is scattered~at~different granularity levels, unlike the well-documented~and~structured code of ML models and rules. 
In detail, data processing code includes data processing components (e.g., \textit{PolicyEnsemble}), general data processing classes and functions in \textit{Shared} module, and specific data processing snippets in components entangled with model or rule usages. 
On the one hand, it could become troublesome to identify and understand the semantics of all data processing code for application developers. A specific example is that data pre-processing code also exists in model definition class of \texttt{TransformerRasaModel}, including \textit{Formant Conversion} and \textit{Data Batch} code. It could be explicitly helpful to automatically extract and analyze the semantics of data processing code  with techniques like program analysis \cite{pycg}. On the other hand, it would be challenging to maintain and test data processing code for system developers, possibly resulting in severe consequences with ML development paradigm shift from model-centric  to data-centric \cite{liangAdvancesChallengesOpportunities2022}. 
In general, building a taxonomy of data processing code would be helpful for the maintaining and testing of data processing code.

\textbf{Code composition patterns.} These non-sequential composition patterns could introduce additional dynamic complexity for ML-enabled systems, e.g., it is too expensive to capture all possible run-time compositions of code snippets~with~static analysis. Although dynamic testing is widely adopted~to~complement the limitations of static analysis in traditional software \cite{fairley1978tutorial}, most existing testing techniques tailored for ML only target at ML model level \cite{ml_testing}. It would be beneficial to extend them to include data processing code and composition patterns.


\section{RQ4: Testing Practice Analysis}
\label{sec:test_case}

\subsection{Methodology}

To answer \textbf{RQ4}, test cases were inspected in three steps.

\textbf{Step 1: Label test cases.} We manually labeled the granularity level, oracle type and ML stage of each test case.
There are three different granularity levels of test cases: (1) Method-level: testing single or multiple methods; (2) Component-level: testing the complete process of a component in training~or~inference stage; (3) Integration-level: testing the current component with upstream components. 
There are four test oracle types: (1) Given input-output pairs: the input and expected output data are given; (2) Component-specific constraints: the constraints must be satisfied according to the implementation of a component, i.e., the sum of confidence scores of the intent list generated by classifiers should equal to 1; (3) Differential executions:~outputs of executions under different settings should change or remain the same, i.e., given the same input, the outputs of an original ML model and its loaded version from disk should remain the same; (4) Exception: whether or not the test case throws exceptions for certain inputs and configurations. 
Finally, the ML stages covered by test cases include training, inference and evaluation stages.
To test the training stage of a component, test cases must first train it and check whether any test oracle is violated.
Note that there may exist several oracle types and ML stages but only one granularity level for each test case in Rasa.
Two of the authors labeled test cases independently, and the third author was involved to resolve disagreements. The Cohen's Kappa coefficient of granularity level, test oracle and ML stage is 0.907, 0.908, and 0.854, respectively. 

\textbf{Step 2: Collect code coverage of test cases.} We collected the statement coverage and branch coverage of code via \textit{pytest-cov}, because Rasa uses \textit{pytest} to run test cases. 

\textbf{Step 3: Collect interaction pattern coverage of test~cases.} We injected logging statements into methods of every component, and then executed test cases to collect the co-executed component sets of each test case. 
All component interaction instances, except \textit{Model Inheritance} instances and some \textit{Usage Constraints} instances that cannot be used together, were tried to be matched with these component sets. 
The matched interaction patterns and instances were considered as covered by test cases.

\subsection{Results}

  \begin{table*}[ht]
        \begin{center}
        \caption{Code Coverage and Labeled Statistics of Test Cases}
        \vspace{-5pt}
        \begin{tabular}{cccccccccccccc}
          \hline
          \multirow{2}{*}{\textbf{Module}} &
            \multirow{2}{*}{\textbf{Total}} &
            \multicolumn{3}{c}{\textbf{Test Case Type}} &
            \multicolumn{3}{c}{\textbf{Test Case Stage}} &
            \multicolumn{4}{c}{\textbf{Oracle Type}} &
            \multicolumn{2}{c}{\textbf{Code Coverage}} \\ \cline{3-14} 
           &
             &
            Meth. &
            Comp. &
            Integ. &
            Infer. &
            Train &
            Eval. &
            I-O &
            C-S &
            Diff. &
            Exception &
            Stat. Cov. &
            Bran. Cov. \\ \hline
          \multicolumn{1}{c|}{Tokenizer} &
            \multicolumn{1}{c|}{27} &
            7 &
            20 &
            \multicolumn{1}{c|}{0} &
            25 &
            14 &
            \multicolumn{1}{c|}{0} &
            24 &
            1 &
            0 &
            \multicolumn{1}{c|}{3} &
            97.4\% &
            96.8\% \\
          \multicolumn{1}{c|}{Featurizer} &
            \multicolumn{1}{c|}{62} &
            13 &
            14 &
            \multicolumn{1}{c|}{35} &
            56 &
            40 &
            \multicolumn{1}{c|}{0} &
            46 &
            5 &
            3 &
            \multicolumn{1}{c|}{8} &
            95.7\% &
            94.9\% \\
          \multicolumn{1}{c|}{IntentClassifier} &
            \multicolumn{1}{c|}{36} &
            7 &
            15 &
            \multicolumn{1}{c|}{14} &
            29 &
            30 &
            \multicolumn{1}{c|}{0} &
            18 &
            11 &
            7 &
            \multicolumn{1}{c|}{1} &
            92.5\% &
            89.5\% \\
          \multicolumn{1}{c|}{EntityExtractor} &
            \multicolumn{1}{c|}{41} &
            6 &
            19 &
            \multicolumn{1}{c|}{16} &
            34 &
            31 &
            \multicolumn{1}{c|}{0} &
            18 &
            14 &
            8 &
            \multicolumn{1}{c|}{1} &
            92.3\% &
            90.1\% \\
          \multicolumn{1}{c|}{Selector} &
            \multicolumn{1}{c|}{13} &
            4 &
            6 &
            \multicolumn{1}{c|}{3} &
            9 &
            12 &
            \multicolumn{1}{c|}{0} &
            6 &
            5 &
            1 &
            \multicolumn{1}{c|}{1} &
            68.3\% &
            66.4\% \\
          \multicolumn{1}{c|}{Policy} &
            \multicolumn{1}{c|}{165} &
            77 &
            88 &
            \multicolumn{1}{c|}{0} &
            105 &
            127 &
            \multicolumn{1}{c|}{0} &
            117 &
            51 &
            0 &
            \multicolumn{1}{c|}{20} &
            95.7\% &
            94.5\% \\
          \multicolumn{1}{c|}{Shared} &
            \multicolumn{1}{c|}{138} &
            129 &
            2 &
            \multicolumn{1}{c|}{7} &
            90 &
            84 &
            \multicolumn{1}{c|}{0} &
            89 &
            42 &
            2 &
            \multicolumn{1}{c|}{16} &
            92.3\% &
            91.4\% \\
          \multicolumn{1}{c|}{Total} &
            \multicolumn{1}{c|}{461} &
            240 &
            156 &
            \multicolumn{1}{c|}{65} &
            331 &
            317 &
            \multicolumn{1}{c|}{47} &
            312 &
            123 &
            15 &
            \multicolumn{1}{c|}{49} &
            93.2\% &
            92.0\% \\ \hline
          \end{tabular}
        \label{test_case_statistics}
        \end{center}
        \end{table*}

The code coverage and labeled statistics of test cases are shown in Table \ref{test_case_statistics}. 
(1) The total statement coverage and branch coverage of code reach 93.2\% and 92.0\%, which~is~much higher than 21.5\% and 13.3\% in Apollo \cite{pengFirstLookIntegration2020}. 
(2) The coverage of \textit{Selector} is only 68.3\% and 66.4\%, because \textit{Selector} has two candidate ML models while only one of them was tested. 
(3) There are 240 (52.0\%) method-level, 156 (33.8\%) component-level and 65 (14.1\%) integration-level test cases. 
(4) There is no integration-level test cases in \textit{Policy}, because \textit{Policy} was tested with given intents and entities input from developers without the dependency of NLU modules. 
(5) Inference and training stages have similar test case quantities.
(6) Only test cases in \textit{Shared} module cover evaluation stage, because \textit{Shared} module provides the evaluation code for all components.
(7) There are 312 (67.7\%), 123 (26.3\%), 15 (3.3\%), and 49 (10.6\%)~test cases with given input-output pairs, component-specific constraints, differential executions and exception test oracles.

        \begin{table}[!t]
            \caption{Test Coverage of Component Interactions}
            \vspace{-10pt}
            \begin{center}
            \tabcolsep=1.0mm
            \begin{tabular}{llll}
            \toprule
            \textbf{Category}
            & \textbf{Sub-Category}
            & \textbf{Cov. Patterns}
            & \textbf{Cov. Instances} \\
            \midrule
            \multirow{5}*{Inter-module}
            &Data Dependency &9/25 &17/95  \\
            &Confidence Checking &0/2 &0/50  \\
            &Output Selection &0/1 &0/18    \\
            &Output Refinement &1/1 & 1/4   \\
            &Usage Constraints &3/3 &3/3    \\
            \midrule
            \multirow{2}*{Intra-module}
            &Functionality Equivalence &2/2 &3/18  \\
            &Prioriy Order &1/1 &4/7  \\
            &Usage Constraints &2/2 &2/4 \\
            \midrule
            Total
            &  &18/37 &30/199 \\
            \bottomrule
            \end{tabular}		
            \label{component_interaction_coverage}
            \end{center}
        \end{table}

        As Table \ref{component_interaction_coverage} shows, the test coverage of interactions~is~relatively low, i.e., 18 (48.6\%) of 37 patterns and 30 (15.1\%)~of~199 instances are covered. This is because only integration-level tests cover components interactions.
        In particular, \textit{Confidence Checking} and \textit{Output Selection} are not covered.
        
        \subsection{Implications} 
        
        \textbf{Low test coverage of component interactions.} It is difficult to achieve a high test coverage of component interactions,~due to the complexity caused by huge configuration space~and~hidden interactions. The only test cases that cover component~interactions (i.e., integration-level test cases) contribute no more than 15\% test cases.
        Yet, integration-level test cases~can~cover~and kill more mutants than component-level and method-level test cases, as many mutants do not manifest~in~non-integration-level test cases \cite{integration_test}.
        Therefore, it is crucial to generate integration-level test cases for ML-enabled systems.
        
        
        \textbf{Limited test oracle types.} 
        It is challenging and time-consuming to write test cases with given input-output pairs and component-specific constraints oracles, due to the complexity brought by lacking of specification for interactions. 
        As~a~result, those test cases without the need of specification of interactions, that is, differential executions and exception test oracles, have been widely utilized to tackle the oracle problem~in~test~case generation techniques for traditional software, such as differential testing \cite{evans2007differential}, fuzzing \cite{liang2018fuzzing} and search-based testing \cite{mcminn2011search}.
        Besides, we find that test cases with these two oracles have a similar capability to kill mutants similar to component-specific constraints oracle (see \textbf{RQ5}). In spite of this, only 13.9\% test cases in Rasa are written with differential executions and exception test oracles, implying that there could be a big room to apply these two test oracle types in test case generation techniques for ML-enabled systems.


\section{RQ5: Mutation Testing Analysis}
\subsection{Methodology}

To answer \textbf{RQ5}, we performed an analysis of mutation~testing~\cite{mutation_survey}. 
It applies mutators to generate versions of faulty~code, i.e., mutants. For every mutant, test cases were executed~to~collect the testing results to decide whether~the~mutant~was~killed. 
As Rasa contains both ML components and rule-based components, we considered both mutators for traditional software (i.e., syntactic mutators) and ML-specific mutators.
As Table~\ref{mutation_operator} shows, we used 9 syntactic mutators from Jia et al. \cite{JiaMutation} and 11 ML specific mutators from DeepCrime \cite{DeepCrime}. 

We list steps of mutation analysis in the following.

\textbf{Step 1: Generate mutants.}
We generated syntactic mutants using \textit{mutmut} \cite{mutmut}. 
We used two groups of syntactic~mutators, i.e., \textit{Logic} and \textit{Value}, which mutate the logic flow~and~variable value. 
Besides, we generated ML specific mutants with DeepCrime \cite{DeepCrime}. 
We used 4 of 8 mutation groups in DeepCrime (\textit{Activation}, \textit{Regularization}, \textit{Weights} and \textit{Optimization}). For others, mutators in \textit{Training Data} and \textit{Validation} groups are not the focus of this paper; \textit{Hyperparameters} group is not included, as hyperparameters in Rasa are specified with configuration files by developers; and \textit{Loss Function} group is not applicable,~as the loss functions in Rasa are all implemented from scratch, while the mutators provided by DeepCrime are only to replace the Keras loss function API with another one. Besides, we only generated mutants for 6 labeled code cateogries in \textbf{RQ3}, excluding general utils code. We generated no more than 30 mutants for every Python class to reduce~potential bias. We also only modified one AST node for every mutant.

\textbf{Step 2: Perform mutation testing analysis with test cases.} For every mutant, only test cases that cover the mutated line were collected (from test coverage data in \textbf{RQ4}) and executed to save running time. 
If any test case fails on a mutant, the mutant is considered as killed by the test case. Otherwise, the mutant is considered as survived. 
A test case could fail~with three symptoms: (1) an assertion fails; (2) an execution or runtime error manifests; and (3) the test case times out. 
The maximum time for a test case to run is 10 times of its running time in the original clean code version. 
Besides, test cases were executed 3 times for every mutant to avoid flaky tests. We found that all test case statuses remain same for three runs. 

\textbf{Step 3: Perform mutation testing analysis with test~data.} For those survived mutants in Step 2, we assessed the impact of them with 3 default configuration files and the restaurant domain data in \textit{Multiwoz} \cite{multiwoz}, which is a widely used multi-domain dataset to evaluate the performance of TDS.
Given a configuration file, only components specified in it are included in the Rasa pipeline, thus mutated nodes of some survived mutants from Step 2 will not be executed~as~they~are~not~\textit{impacted} by the configuration.
Due to the stochastic nature of machine learning programs, we trained both the mutated program and original program for 5 times with 80/20 data split into train/test data randomly, and decided whether the performance metrics of two versions are statistically significant with non-negligible and non-small effect size.
We followed the same formula to decide whether a mutant is killed with the test data as \cite{mutation_evaluation, DeepCrime}, with the threshold of significance value is 0.05 and of effect size is 0.5.
We adopted F1 scores of \textit{IntentClassifier}, \textit{EntityExtractor} and \textit{Policy} as performance metrics, i.e., if the F1 score in any of the three modules is statistically different between two code versions, the mutant is marked as killed by test data.
\subsection{Results}


\begin{table}[]
  \tabcolsep=1.0mm
  \begin{center}
  \caption{Mutation Testing Results}
  \vspace{-5pt}
  \begin{tabular}{ccccccc}
  \hline
  \multirow{2}{*}{\textbf{Mutation Group}} & \multirow{2}{*}{\textbf{Operator}} & \multirow{2}{*}{\textbf{Total}} & \multicolumn{2}{c}{\textbf{Test Case}} & \multicolumn{2}{c}{\textbf{Test Data}} \\ \cline{4-7} 
                                                       &                            &                       & Killed & Survived              & Impacted & Killed  \\ \hline
  \multicolumn{1}{c|}{\multirow{6}{*}{Logic}}          & \multicolumn{1}{c|}{ArOR}  & \multicolumn{1}{c|}{109} &86        & \multicolumn{1}{c|}{23} &10          & 1 \\
  \multicolumn{1}{c|}{}                                & \multicolumn{1}{c|}{ComOR} & \multicolumn{1}{c|}{109} &88        & \multicolumn{1}{c|}{21} &6          &0         \\
  \multicolumn{1}{c|}{}                                & \multicolumn{1}{c|}{LogOR} & \multicolumn{1}{c|}{145} &112        & \multicolumn{1}{c|}{33} &14          &0         \\
  \multicolumn{1}{c|}{}                                & \multicolumn{1}{c|}{AsOR}  & \multicolumn{1}{c|}{20} &19        & \multicolumn{1}{c|}{1} &0          &0         \\
  \multicolumn{1}{c|}{}                                & \multicolumn{1}{c|}{MemOR} & \multicolumn{1}{c|}{32} &30        & \multicolumn{1}{c|}{2} &0          &0         \\
  \multicolumn{1}{c|}{}                                & \multicolumn{1}{c|}{KVR}   & \multicolumn{1}{c|}{12} &7        & \multicolumn{1}{c|}{5} &1          &0         \\ \hline
  \multicolumn{1}{c|}{\multirow{3}{*}{Value}}          & \multicolumn{1}{c|}{BVR}   & \multicolumn{1}{c|}{64} &32        & \multicolumn{1}{c|}{32} &9          &0         \\
  \multicolumn{1}{c|}{}                                & \multicolumn{1}{c|}{NVR}   & \multicolumn{1}{c|}{224} &180        & \multicolumn{1}{c|}{64} &18          &2         \\
  \multicolumn{1}{c|}{}                                & \multicolumn{1}{c|}{AsVR}  & \multicolumn{1}{c|}{582} &525        & \multicolumn{1}{c|}{67} &10          &0         \\ \hline
  \multicolumn{1}{c|}{\multirow{3}{*}{Activation}}     & \multicolumn{1}{c|}{ACH}   & \multicolumn{1}{c|}{22} &3        & \multicolumn{1}{c|}{19} &18          &6         \\
  \multicolumn{1}{c|}{}                                & \multicolumn{1}{c|}{ARM}   & \multicolumn{1}{c|}{2} &0        & \multicolumn{1}{c|}{2} &1          &0         \\
  \multicolumn{1}{c|}{}                                & \multicolumn{1}{c|}{AAL}   & \multicolumn{1}{c|}{22} &11        & \multicolumn{1}{c|}{11} &11          &2         \\ \hline
  \multicolumn{1}{c|}{\multirow{3}{*}{Regularization}} & \multicolumn{1}{c|}{RAW}   & \multicolumn{1}{c|}{6} &0        & \multicolumn{1}{c|}{6} &3          &3         \\
  \multicolumn{1}{c|}{}                                & \multicolumn{1}{c|}{RCW}   & \multicolumn{1}{c|}{10} &0        & \multicolumn{1}{c|}{10} &10          &0         \\
  \multicolumn{1}{c|}{}                                & \multicolumn{1}{c|}{RRW}   & \multicolumn{1}{c|}{5} &0        & \multicolumn{1}{c|}{5} &5          &0         \\ \hline
  \multicolumn{1}{c|}{\multirow{3}{*}{Weights}}        & \multicolumn{1}{c|}{WCI}   & \multicolumn{1}{c|}{24} &10        & \multicolumn{1}{c|}{14} &13          &1         \\
  \multicolumn{1}{c|}{}                                & \multicolumn{1}{c|}{WAB}   & \multicolumn{1}{c|}{4} &0        & \multicolumn{1}{c|}{4} &3          &1         \\
  \multicolumn{1}{c|}{}                                & \multicolumn{1}{c|}{WRB}   & \multicolumn{1}{c|}{3} &1        & \multicolumn{1}{c|}{2} &2          &0         \\ \hline
  \multicolumn{1}{c|}{\multirow{2}{*}{Optimization}}   & \multicolumn{1}{c|}{OCH}   & \multicolumn{1}{c|}{24} &2        & \multicolumn{1}{c|}{22} &9          &6         \\
  \multicolumn{1}{c|}{}                                & \multicolumn{1}{c|}{OCG}   & \multicolumn{1}{c|}{8} &0        & \multicolumn{1}{c|}{8} &3          &0         \\ \hline
  \multicolumn{1}{c|}{\multirow{1}{*}{Total}}   & \multicolumn{1}{c|}{}   & \multicolumn{1}{c|}{1447} &1106        & \multicolumn{1}{c|}{341} &146          &22         \\ \hline
  \end{tabular}
  \label{mutation_operator}
  \end{center}
  \end{table}
  

  The mutation testing results by each mutator are shown~in Table \ref{mutation_operator}. There are 1447 mutants generated, 1106 (76.4\%)~mutants killed by test cases, 341 (23.6\%) mutants survived, 146 (10.1\%) mutants impacted, and 22 (1.5\%) mutants killed by test data. 
  Only 146 mutants from 341 survived mutants impact the default 3 Rasa pipelines, which shows that the huge configuration space is challenging to be tested  adequately.
  81.3\% syntactic mutants and 20.0\% ML specific mutants are killed by test cases, while 4.4\% syntactic mutants and 24.4\% ML specific mutants from impacted mutants are killed by test data.
  It shows that test case is much more effective to detect syntactic mutants and slightly less effective to detect ML specific mutants than test data. 
  The killed syntactic mutants and ML-specific mutants by test data cause the F1 score degradation of \textit{IntentClassifier}, \textit{EntityExtractor} and \textit{Policy} by 20.8\%, 0.8\%, 3.6\% and 11.1\%, 13.4\%, 5.7\% on average.

  
  

  \begin{table}[!t]
  \begin{center}
  \caption{Mutant Location Results}
  \vspace{-5pt}
  \scalebox{1}{
    \begin{tabular}{cccccc}
      \hline
      \multirow{2}{*}{\textbf{Location}} & \multirow{2}{*}{\textbf{Total}} & \multicolumn{2}{c}{\textbf{Test Case Result}} & \multicolumn{2}{c}{\textbf{Test Data Result}} \\ \cline{3-6} 
                                       &                          & Killed & Survived                 & Impacted & Killed \\ \hline
      \multicolumn{1}{c|}{Data Prep.}  & \multicolumn{1}{c|}{385} & 326    & \multicolumn{1}{c|}{59}  & 23       & 0      \\
      \multicolumn{1}{c|}{Data Post.}  & \multicolumn{1}{c|}{271} & 222    & \multicolumn{1}{c|}{49}  & 5        & 0      \\
      \multicolumn{1}{c|}{Model Usage} & \multicolumn{1}{c|}{307} & 243    & \multicolumn{1}{c|}{64}  & 4        & 0      \\
      \multicolumn{1}{c|}{Model Def.}  & \multicolumn{1}{c|}{364} & 224    & \multicolumn{1}{c|}{140} & 99       & 22     \\
      \multicolumn{1}{c|}{Rule Usage}  & \multicolumn{1}{c|}{4}   & 4      & \multicolumn{1}{c|}{0}   & 0        & 0      \\
      \multicolumn{1}{c|}{Rule Def.}   & \multicolumn{1}{c|}{115} & 101    & \multicolumn{1}{c|}{14}  & 4        & 0      \\ \hline
      \end{tabular}}
    \label{mutation_location}
    \end{center}
    \end{table}

  The mutation testing results w.r.t. the location of mutants are shown in Table \ref{mutation_location}. 224 (61.5\%) of 364 mutants in model definition code, and 896 (82.8\%) of 1082 mutants in other code categories are killed by test cases. 
  In particular, few mutants in code categories except model definition are impacted and killed by test data, which implies that test data is only effective to kill mutants in model definition code.

    \begin{table}[!t]
        \begin{center}
          \tabcolsep=1.0mm
            \caption{Test Case Mutation Results}
            \vspace{-5pt}
        \scalebox{0.95}{
        \begin{tabular}{c|c|cccc}
        \hline
        \textbf{Category} & \textbf{Type} &\textbf{Test Num.} &\textbf{Strong Test Num.}     & \textbf{Covered}        &\textbf{Killed}  \\ \hline
        \multirow{3}{*}{Granularity}   & Method &240 &59 &947     & 635      \\
                                          & Component &156 &31 &1121       &709     \\
                                          & Integration &65 &29 &903    &613      \\ \hline
        \multirow{4}{*}{Stage}  & Infer. &331 &86 &1358           &995      \\
                                          & Training &317 &75 &1184        &847      \\
                                          & Evaluation &47 &11 &772 &476       \\ \hline
        \multirow{4}{*}{Oracle Type}  & I-O &312 &98 &1298    &956      \\
                                          & C-S  &123 &19 &1103  &707      \\
                                          & Diff &15 &3  &625   &338    \\
                 & Exception &49  &6 &686 &352   \\ \hline
        \end{tabular}}
        \label{test_mutation}
    \end{center}
        \end{table}
    
    We investigated the capability to detect mutants~w.r.t.~different categories of test cases, by calculating the ratio of \textit{strong test case number} to \textit{all test case number}, and the ratio of \textit{killed mutants} to \textit{covered mutants} of them.  
    We define \textit{strong test case} as the test case that kills equal or more than 75\% of its covered mutants. 
    As Table \ref{test_mutation} shows, test cases in integration level have the highest ratio of strong test case (44.6\%) and highest ratio of killed mutants (67.9\%) among three granularity levels.
    Test cases with given input-output test oracle have the highest ratio of strong test case (31.4\%)  and highest ratio of killed mutants (73.7\%) among four oracle types, while test cases with other three oracle types have similar ratios.

\subsection{Implications}

\textbf{Non-ML specific bugs and test cases in ML-enabled~systems.} 
Complexity from data processing code causes that~non-ML specific bugs are prone to be introduced.
Compared~with test data, test case is more effective to detect syntactic mutants, i.e., non-ML specific bugs.
Moreover, it is notorious~for~developers to analyze, localize and fix bugs in ML programs~according to test data, thus interpreting \cite{interpretability}, debugging \cite{abid2022meaningfully} and repairing \cite{sun2022causality} techniques have been developed for ML models. 
It is easier for developers to localize and fix bugs with failed test cases by analyzing violated test oracles.
Thus, we claim that non-ML specific bugs and test cases in  ML-enbaled systems should be paid more attention to.
Although there~is~a~rich~set of test cases in Rasa that achieve high code coverage, the kill ratio of mutants remains to be improved (76.4\%), especially of ML-specific mutants (29.8\%).
The applicability and limitations of existing test case generation, selection and quality assurance techniques in ML-enabled systems are worthwhile to be explored \cite{kazmi2017effective, di2013coverage}. 

%

\textbf{Challenges of test data to kill mutants.}
Existing researches on mutation testing for ML programs only evaluated mutants with test data to decide whether they can be killed~\cite{DeepCrime,DeepMutation++,DeepMutation,mutation_evaluation,JiaMutation}.
However, the capability of test data to kill mutants in large-scale ML-enabled systems is limited for two reasons.
First, due to complexity from configurations, only part of mutants will impact the components of actual configured systems.
Second, the amount and distributions of training data and test data affect the results a lot.
For example, we tried to train the clean code version and mutated version with 75\% of original training data, the number of killed mutants changed from 22 to 83, which means some bugs may only manifest under specific training data settings. 
Therefore, system developers should evaluate and test ML-enabled systems under more possible configurations and data settings that may be used by application developers to detect potential bugs.


\section{threats}

First, our study conducts a case study on Rasa, a widely~used task-oriented industrial dialogue system. It is not clear whether our results can be generalized to other ML-enabled systems. However, we believe it is a good start to take a system view~for ML-enabled systems. Second, our study involves~a~lot~of~manual analyses of Rasa source code and documentations, which may incur biases. To reduce them, two of the authors conduct manual analysis separately, and a third author is involved to resolve disagreements. Third,~the~mutators~that we adopt~may not simulate real-world bugs. To mitigate it,  we decide~to~use mutators from DeepCrime \cite{DeepCrime}, whose mutators are actually summarized from real word ML bugs. 

\section{Related Work}

\textbf{Study of ML-Enabled Systems}. While much of the attention has been on ML models, less attention has been paid on system-level analysis~\cite{Christian2022}. Peng et al.~\cite{pengFirstLookIntegration2020} investigated~the~integration of ML models in Apollo by analyzing how ML models~interact with the system and how is the current testing effort. Besides, Nahar et al.~\cite{Nahar2022} explored collaboration challenges between data scientists and software engineers through interviews.~Amershi et al.~\cite{Amershi2019} and Bernardi et al.~\cite{Bernardi2019} reported~challenges and~practices of MLOps (from model requirement~to~model~monitoring) at Microsoft and Booking.com. Although they~still~take~a~model-centric view,~they~emphasize~that~models can~be~complexly~entangled to cause non-monotonic~errors~\cite{Amershi2019} and model quality~improvement does not necessarily indicate system value~gain~\cite{Bernardi2019}. Further, Yokoyama~\cite{Yokoyama2019} developed an architectural pattern to separate ML and non-ML components, while Serban~and~Visser \cite{Serban2022} surveyed architectural challenges for ML-enabled systems. Sculley et al.~\cite{hidden_technical_debt} identified ML-specific technical debt~in~ML-enabled systems, while Tang et al.~\cite{tang2021empirical} further derived new~ones from real-world code refactorings. In addition, some attempts were made on the problem of ML component entanglement~\cite{Amershi2019}, e.g., performing metamorphic testing on a system with two~ML components~\cite{Zhang2016}, troubleshooting failures in a system with~three ML components by human intellect~\cite{nushi2017human}, and decomposing~errors in a system with two or three ML components~\cite{fix_that_fails}. These studies explore the interaction among models but only~on~simple systems. Moreover, Abdessalem et al.~\cite{Abdessalem2018, Abdessalem2020} studied the feature interaction failures in self-driving systems, and proposed testing and repairing approaches to automatically detect and fix them. Apel et al.~\cite{feature_interaction} also discussed feature interactions in ML-enabled systems, and suggested strategies to cope with them.

The main difference from the previous work is that we~take~a large-scale complex ML-enabled system, explore its complexity at three levels, and analyze the impact of its complexity~on~testing. The closest work is Peng et al.'s~\cite{pengFirstLookIntegration2020}, but we report~a~deeper complexity analysis and also conduct a testing impact analysis.

\textbf{Mutation Testing for DL Models}. Jia et al.~\cite{JiaMutation}~used~syntactic mutators for traditional programs to DL models.~DeepMutation \cite{DeepMutation} and DeepMutation++~\cite{DeepMutation++} defined DL-specific mutators. DeepCrime~\cite{DeepCrime} derived DL-specific  mutators based~on real~DL bugs. 
Jahangirova and Tonella~\cite{mutation_evaluation} evaluated syntactic and DL-specific mutators. These studies are focused on model-level mutation, while we target at system-level mutation.

\textbf{Testing for Dialogue Systems}. Bozic and Wotawa~\cite{Bozic2018}~proposed a security testing approach for chatbots to prevent cross-site scripting and SQL injection. Bozic et al.~\cite{Bozic2019a}~tested a hotel booking chatbot via planning. Bozic and Wotawa~\cite{Bozic2019b}~introduced a metamorphic testing approach for chatbots. Similarly, Liu et al.~\cite{liu2021dialtest} used semantic metamorphic relations to test the NLU module in dialogue systems. Despite the effort,~less~attention has been paid on system-level testing of dialogue systems.


\section{conclusion}


We present a comprehensive study on Rasa to characterize its complexity at three levels and the impact of its complexity on testing from two perspectives.
Furthermore, we highlight practical implications to improve software engieering for ML-enabled systems.
All study data and source code used in this paper are available at \url{https://rasasystemcomplexity.github.io/}.


{\footnotesize
\bibliographystyle{IEEEtranS}
\bibliography{IEEEabrv, src/reference}

\begin{thebibliography}{10}
\providecommand{\url}[1]{#1}
\csname url@samestyle\endcsname
\providecommand{\newblock}{\relax}
\providecommand{\bibinfo}[2]{#2}
\providecommand{\BIBentrySTDinterwordspacing}{\spaceskip=0pt\relax}
\providecommand{\BIBentryALTinterwordstretchfactor}{4}
\providecommand{\BIBentryALTinterwordspacing}{\spaceskip=\fontdimen2\font plus
\BIBentryALTinterwordstretchfactor\fontdimen3\font minus
  \fontdimen4\font\relax}
\providecommand{\BIBforeignlanguage}[2]{{%
\expandafter\ifx\csname l@#1\endcsname\relax
\typeout{** WARNING: IEEEtranS.bst: No hyphenation pattern has been}%
\typeout{** loaded for the language `#1'. Using the pattern for}%
\typeout{** the default language instead.}%
\else
\language=\csname l@#1\endcsname
\fi
#2}}
\providecommand{\BIBdecl}{\relax}
\BIBdecl

\bibitem{Abdessalem2018}
R.~B. Abdessalem, A.~Panichella, S.~Nejati, L.~C. Briand, and T.~Stifter,
  ``Testing autonomous cars for feature interaction failures using
  many-objective search,'' in \emph{Proceedings of the 33rd ACM/IEEE
  International Conference on Automated Software Engineering}, 2018, p.
  143–154.

\bibitem{Abdessalem2020}
------, ``Automated repair of feature interaction failures in automated driving
  systems,'' in \emph{Proceedings of the 29th ACM SIGSOFT International
  Symposium on Software Testing and Analysis}, 2020, pp. 88--100.

\bibitem{abid2022meaningfully}
A.~Abid, M.~Yuksekgonul, and J.~Zou, ``Meaningfully debugging model mistakes
  using conceptual counterfactual explanations,'' in \emph{Proceedings of the
  International Conference on Machine Learning}, 2022, pp. 66--88.

\bibitem{Aggarwal2019}
A.~Aggarwal, P.~Lohia, S.~Nagar, K.~Dey, and D.~Saha, ``Black box fairness
  testing of machine learning models,'' in \emph{Proceedings of the 2019 27th
  ACM Joint Meeting on European Software Engineering Conference and Symposium
  on the Foundations of Software Engineering}, 2019, p. 625–635.

\bibitem{Amershi2019}
S.~Amershi, A.~Begel, C.~Bird, R.~DeLine, H.~Gall, E.~Kamar, N.~Nagappan,
  B.~Nushi, and T.~Zimmermann, ``Software engineering for machine learning: A
  case study,'' in \emph{Proceedings of the 41st International Conference on
  Software Engineering: Software Engineering in Practice}, 2019, pp. 291--300.

\bibitem{dask}
\BIBentryALTinterwordspacing
Anaconda. (2022) Dask. [Online]. Available:
  \url{https://docs.dask.org/en/stable/}
\BIBentrySTDinterwordspacing

\bibitem{website}
\BIBentryALTinterwordspacing
Anonymous. (2022) Understanding the complexity and its impact on testing in
  ml-enabled systems. [Online]. Available:
  \url{https://rasasystemcomplexity.github.io/}
\BIBentrySTDinterwordspacing

\bibitem{feature_interaction}
S.~Apel, C.~Kästner, and E.~Kang, ``Feature interactions on steroids: On the
  composition of ml models,'' \emph{IEEE Software}, vol.~39, no.~3, pp.
  120--124, 2022.

\bibitem{Baluta2021}
T.~Baluta, Z.~L. Chua, K.~S. Meel, and P.~Saxena, ``Scalable quantitative
  verification for deep neural networks,'' in \emph{Proceedings of the 43rd
  International Conference on Software Engineering: Companion Proceedings},
  2021, p. 248–249.

\bibitem{Bernardi2019}
L.~Bernardi, T.~Mavridis, and P.~Estevez, ``150 successful machine learning
  models: 6 lessons learned at booking.com,'' in \emph{Proceedings of the 25th
  ACM SIGKDD International Conference on Knowledge Discovery \& Data Mining},
  2019, p. 1743–1751.

\bibitem{rasa}
T.~Bocklisch, J.~Faulkner, N.~Pawlowski, and A.~Nichol, ``Rasa: Open source
  language understanding and dialogue management,'' \emph{CoRR}, vol.
  abs/1712.05181, 2017.

\bibitem{Bozic2019a}
J.~Bozic, O.~A. Tazl, and F.~Wotawa, ``Chatbot testing using ai planning,'' in
  \emph{Proceedings of the IEEE International Conference On Artificial
  Intelligence Testing}, 2019, pp. 37--44.

\bibitem{Bozic2018}
J.~Bozic and F.~Wotawa, ``Security testing for chatbots,'' in \emph{Proceedings
  of the IFIP International Conference on Testing Software and Systems}, 2018,
  pp. 33--38.

\bibitem{Bozic2019b}
------, ``Testing chatbots using metamorphic relations,'' in \emph{Proceedings
  of the IFIP International Conference on Testing Software and Systems}, 2019,
  pp. 41--55.

\bibitem{multiwoz}
P.~Budzianowski, T.-H. Wen, B.-H. Tseng, I.~Casanueva, S.~Ultes, O.~Ramadan,
  and M.~Ga{\v{s}}i{\'c}, ``Multiwoz - a large-scale multi-domain wizard-of-oz
  dataset for task-oriented dialogue modelling,'' in \emph{Proceedings of the
  Conference on Empirical Methods in Natural Language Processing}, 2018, pp.
  5016--5026.

\bibitem{chaudhuri-etal-2018-improving}
D.~Chaudhuri, A.~Kristiadi, J.~Lehmann, and A.~Fischer, ``Improving response
  selection in multi-turn dialogue systems by incorporating domain knowledge,''
  in \emph{Proceedings of the 22nd Conference on Computational Natural Language
  Learning}, 2018, pp. 497--507.

\bibitem{devlin2018bert}
J.~Devlin, M.-W. Chang, K.~Lee, and K.~Toutanova, ``Bert: Pre-training of deep
  bidirectional transformers for language understanding,'' \emph{arXiv preprint
  arXiv:1810.04805}, 2018.

\bibitem{di2013coverage}
D.~Di~Nardo, N.~Alshahwan, L.~Briand, and Y.~Labiche, ``Coverage-based test
  case prioritisation: An industrial case study,'' in \emph{2013 IEEE Sixth
  International Conference on Software Testing, Verification and Validation},
  2013, pp. 302--311.

\bibitem{Dola2021}
S.~Dola, M.~B. Dwyer, and M.~L. Soffa, ``Distribution-aware testing of neural
  networks using generative models,'' in \emph{Proceedings of the IEEE/ACM 43rd
  International Conference on Software Engineering}, 2021, pp. 226--237.

\bibitem{eddy1996hidden}
S.~R. Eddy, ``Hidden markov models,'' \emph{Current opinion in structural
  biology}, vol.~6, no.~3, pp. 361--365, 1996.

\bibitem{evans2007differential}
R.~B. Evans and A.~Savoia, ``Differential testing: a new approach to change
  detection,'' in \emph{The 6th Joint Meeting on European software engineering
  conference and the ACM SIGSOFT Symposium on the Foundations of Software
  Engineering: Companion Papers}, 2007, pp. 549--552.

\bibitem{transformers}
\BIBentryALTinterwordspacing
H.~Face. (2022) Transformers. [Online]. Available:
  \url{https://huggingface.co/docs/transformers/index}
\BIBentrySTDinterwordspacing

\bibitem{duckling}
\BIBentryALTinterwordspacing
Facebook. (2022) Duckling. [Online]. Available:
  \url{https://github.com/facebook/duckling/}
\BIBentrySTDinterwordspacing

\bibitem{fairley1978tutorial}
R.~E. Fairley, ``Tutorial: Static analysis and dynamic testing of computer
  software,'' \emph{Computer}, vol.~11, no.~4, pp. 14--23, 1978.

\bibitem{Feng2020}
Y.~Feng, Q.~Shi, X.~Gao, J.~Wan, C.~Fang, and Z.~Chen, ``Deepgini: Prioritizing
  massive tests to enhance the robustness of deep neural networks,'' in
  \emph{Proceedings of the 29th ACM SIGSOFT International Symposium on Software
  Testing and Analysis}, 2020, p. 177–188.

\bibitem{Foo2019TheDO}
D.~Foo, J.~Yeo, H.~Xiao, and A.~Sharma, ``The dynamics of software composition
  analysis,'' \emph{CoRR}, vol. abs/1909.00973, 2019.

\bibitem{TensorHub}
\BIBentryALTinterwordspacing
Google. (2022) Tensorhub. [Online]. Available:
  \url{https://tensorflow.google.cn/hub}
\BIBentrySTDinterwordspacing

\bibitem{XinHe2021AutoMLAS}
X.~He, K.~Zhao, and X.~Chu, ``Automl: A survey of the state-of-the-art,''
  \emph{Knowledge Based Systems}, vol. 212, 2021.

\bibitem{henderson2019convert}
M.~Henderson, I.~Casanueva, N.~Mrk{\v{s}}i{\'c}, P.-H. Su, T.-H. Wen, and
  I.~Vuli{\'c}, ``Convert: Efficient and accurate conversational
  representations from transformers,'' \emph{CoRR}, 2019.

\bibitem{DeepMutation++}
Q.~Hu, L.~Ma, X.~Xie, B.~Yu, Y.~Liu, and J.~Zhao, ``Deepmutation++: A mutation
  testing framework for deep learning systems,'' in \emph{Proceedings of the
  34th IEEE/ACM International Conference on Automated Software Engineering},
  2019, pp. 1158--1161.

\bibitem{hu2020gpt}
Z.~Hu, Y.~Dong, K.~Wang, K.-W. Chang, and Y.~Sun, ``Gpt-gnn: Generative
  pre-training of graph neural networks,'' in \emph{Proceedings of the 26th ACM
  SIGKDD International Conference on Knowledge Discovery \& Data Mining}, 2020,
  pp. 1857--1867.

\bibitem{dependency_bug}
K.~Huang, B.~Chen, S.~Wu, J.~Cao, L.~Ma, and X.~Peng, ``Demystifying dependency
  bugs in deep learning stack,'' \emph{CoRR}, vol. abs/2207.10347, 2022.

\bibitem{DeepCrime}
N.~Humbatova, G.~Jahangirova, and P.~Tonella, ``Deepcrime: Mutation testing of
  deep learning systems based on real faults,'' in \emph{Proceedings of the
  30th ACM SIGSOFT International Symposium on Software Testing and Analysis},
  2021, p. 67–78.

\bibitem{ibmreport}
\BIBentryALTinterwordspacing
IBM. (2022) Ibm global ai adoption index 2022. [Online]. Available:
  \url{https://www.ibm.com/watson/resources/ai-adoption}
\BIBentrySTDinterwordspacing

\bibitem{mutation_evaluation}
G.~Jahangirova and P.~Tonella, ``An empirical evaluation of mutation operators
  for deep learning systems,'' in \emph{Proceedings of the IEEE 13th
  International Conference on Software Testing, Validation and Verification},
  2020, pp. 74--84.

\bibitem{JiaMutation}
L.~Jia, H.~Zhong, X.~Wang, L.~Huang, and Z.~Li, ``How do injected bugs affect
  deep learning?'' in \emph{Proceedings of the IEEE International Conference on
  Software Analysis, Evolution and Reengineering}, 2022, pp. 793--804.

\bibitem{mutation_survey}
Y.~Jia and M.~Harman, ``An analysis and survey of the development of mutation
  testing,'' \emph{IEEE Transactions on Software Engineering}, vol.~37, no.~5,
  pp. 649--678, 2011.

\bibitem{Christian2022}
\BIBentryALTinterwordspacing
C.~K{\"{a}}stner. (2022) Machine learning in production: From models to
  systems. [Online]. Available:
  \url{https://ckaestne.medium.com/machine-learning-in-production-from-models-to-systems-e1422ec7cd65}
\BIBentrySTDinterwordspacing

\bibitem{kazmi2017effective}
R.~Kazmi, D.~N. Jawawi, R.~Mohamad, and I.~Ghani, ``Effective regression test
  case selection: A systematic literature review,'' \emph{ACM Computing Surveys
  (CSUR)}, vol.~50, no.~2, pp. 1--32, 2017.

\bibitem{Kim2019}
J.~Kim, R.~Feldt, and S.~Yoo, ``Guiding deep learning system testing using
  surprise adequacy,'' in \emph{Proceedings of the 41st International
  Conference on Software Engineering}, 2019, p. 1039–1049.

\bibitem{xlm}
\BIBentryALTinterwordspacing
G.~Lample and A.~Conneau, ``Cross-lingual language model pretraining,'' 2019.
  [Online]. Available: \url{https://arxiv.org/abs/1901.07291}
\BIBentrySTDinterwordspacing

\bibitem{integration_test}
H.~Leung and L.~White, ``A study of integration testing and software regression
  at the integration level,'' in \emph{Proceedings. Conference on Software
  Maintenance 1990}, 1990, pp. 290--301.

\bibitem{Li2020}
Z.~Li, X.~Ma, C.~Xu, J.~Xu, C.~Cao, and J.~L\"{u}, ``Operational calibration:
  Debugging confidence errors for dnns in the field,'' in \emph{Proceedings of
  the 28th ACM Joint Meeting on European Software Engineering Conference and
  Symposium on the Foundations of Software Engineering}, 2020, p. 901–913.

\bibitem{liang2018fuzzing}
H.~Liang, X.~Pei, X.~Jia, W.~Shen, and J.~Zhang, ``Fuzzing: State of the art,''
  \emph{IEEE Transactions on Reliability}, vol.~67, no.~3, pp. 1199--1218,
  2018.

\bibitem{liangAdvancesChallengesOpportunities2022}
W.~Liang, G.~A. Tadesse, D.~Ho, F.-F. Li, M.~Zaharia, C.~Zhang, and J.~Zou,
  ``Advances, challenges and opportunities in creating data for trustworthy
  {AI},'' \emph{Nature Machine Intelligence}, 2022.

\bibitem{liu2019roberta}
Y.~Liu, M.~Ott, N.~Goyal, J.~Du, M.~Joshi, D.~Chen, O.~Levy, M.~Lewis,
  L.~Zettlemoyer, and V.~Stoyanov, ``Roberta: A robustly optimized bert
  pretraining approach,'' \emph{arXiv preprint arXiv:1907.11692}, 2019.

\bibitem{liu2021dialtest}
Z.~Liu, Y.~Feng, and Z.~Chen, ``Dialtest: automated testing for
  recurrent-neural-network-driven dialogue systems,'' in \emph{Proceedings of
  the 30th ACM SIGSOFT International Symposium on Software Testing and
  Analysis}, 2021, pp. 115--126.

\bibitem{DeepMutation}
L.~Ma, F.~Zhang, J.~Sun, M.~Xue, B.~Li, F.~Juefei-Xu, C.~Xie, L.~Li, Y.~Liu,
  J.~Zhao, and Y.~Wang, ``Deepmutation: Mutation testing of deep learning
  systems,'' in \emph{Proceedings of the IEEE 29th International Symposium on
  Software Reliability Engineering}, 2018, pp. 100--111.

\bibitem{Ma2018}
S.~Ma, Y.~Liu, W.-C. Lee, X.~Zhang, and A.~Grama, ``Mode: Automated neural
  network model debugging via state differential analysis and input
  selection,'' in \emph{Proceedings of the 2018 26th ACM Joint Meeting on
  European Software Engineering Conference and Symposium on the Foundations of
  Software Engineering}, 2018, p. 175–186.

\bibitem{mcminn2011search}
P.~McMinn, ``Search-based software testing: Past, present and future,'' in
  \emph{2011 IEEE Fourth International Conference on Software Testing,
  Verification and Validation Workshops}, 2011, pp. 153--163.

\bibitem{mutmut}
\BIBentryALTinterwordspacing
Mutmut. (2022) Mutmut. [Online]. Available:
  \url{https://pypi.org/project/mutmut/}
\BIBentrySTDinterwordspacing

\bibitem{Nahar2022}
N.~Nahar, S.~Zhou, G.~Lewis, and C.~Kästner, ``Collaboration challenges in
  building ml-enabled systems: Communication, documentation, engineering, and
  process,'' in \emph{Proceedings of the IEEE/ACM 44th International Conference
  on Software Engineering}, 2022, pp. 413--425.

\bibitem{nushi2017human}
B.~Nushi, E.~Kamar, E.~Horvitz, and D.~Kossmann, ``On human intellect and
  machine failures: Troubleshooting integrative machine learning systems,'' in
  \emph{Proceedings of the Thirty-First AAAI Conference on Artificial
  Intelligence}, 2017, pp. 1017--1025.

\bibitem{odena19a}
A.~Odena, C.~Olsson, D.~Andersen, and I.~Goodfellow, ``{T}ensor{F}uzz:
  Debugging neural networks with coverage-guided fuzzing,'' in
  \emph{Proceedings of the 36th International Conference on Machine Learning},
  2019, pp. 4901--4911.

\bibitem{Katie2020}
K.~O'Leary and M.~Uchida, ``Common problems with creating machine learning
  pipelines from existing code,'' in \emph{Proceedings of the Third Conference
  on Machine Learning and Systems}, 2020.

\bibitem{gpt2}
\BIBentryALTinterwordspacing
OpenAI. (2022) Gpt2. [Online]. Available:
  \url{https://openai.com/blog/tags/gpt-2/}
\BIBentrySTDinterwordspacing

\bibitem{Paulsen2020b}
B.~Paulsen, J.~Wang, and C.~Wang, ``Reludiff: Differential verification of deep
  neural networks,'' in \emph{Proceedings of the ACM/IEEE 42nd International
  Conference on Software Engineering}, 2020, p. 714–726.

\bibitem{Paulsen2020a}
B.~Paulsen, J.~Wang, J.~Wang, and C.~Wang, ``Neurodiff: Scalable differential
  verification of neural networks using fine-grained approximation,'' in
  \emph{Proceedings of the 35th IEEE/ACM International Conference on Automated
  Software Engineering}, 2020, p. 784–796.

\bibitem{Pei2017}
K.~Pei, Y.~Cao, J.~Yang, and S.~Jana, ``Deepxplore: Automated whitebox testing
  of deep learning systems,'' in \emph{Proceedings of the 26th Symposium on
  Operating Systems Principles}, 2017, p. 1–18.

\bibitem{pengFirstLookIntegration2020}
Z.~Peng, J.~Yang, T.-H.~P. Chen, and L.~Ma, ``A first look at the integration
  of machine learning models in complex autonomous driving systems: A case
  study on apollo,'' in \emph{Proceedings of the 28th ACM Joint Meeting on
  European Software Engineering Conference and Symposium on the Foundations of
  Software Engineering}, pp. 1240--1250.

\bibitem{pycg}
V.~Salis, T.~Sotiropoulos, P.~Louridas, D.~Spinellis, and D.~Mitropoulos,
  ``Pycg: Practical call graph generation in python,'' in \emph{Proceedings of
  the 43rd International Conference on Software Engineering}, 2021, p.
  1646–1657.

\bibitem{Sattler2017LiftingID}
F.~Sattler, A.~von Rhein, T.~Berger, N.~S. Johansson, M.~M. Hard{\o}, and
  S.~Apel, ``Lifting inter-app data-flow analysis to large app sets,''
  \emph{Automated Software Engineering}, vol.~25, no.~2, pp. 315--346, 2017.

\bibitem{hidden_technical_debt}
D.~Sculley, G.~Holt, D.~Golovin, E.~Davydov, T.~Phillips, D.~Ebner,
  V.~Chaudhary, M.~Young, J.-F. Crespo, and D.~Dennison, ``Hidden technical
  debt in machine learning systems,'' in \emph{Proceedings of the 28th
  International Conference on Neural Information Processing Systems}, 2015, p.
  2503–2511.

\bibitem{Serban2022}
A.~Serban and J.~Visser, ``Adapting software architectures to machine learning
  challenges,'' in \emph{Proceedings of the IEEE International Conference on
  Software Analysis, Evolution and Reengineering}, 2022, pp. 152--163.

\bibitem{Shmilovici2005}
A.~Shmilovici.

\bibitem{Singh2019}
G.~Singh, T.~Gehr, M.~P\"{u}schel, and M.~Vechev, ``An abstract domain for
  certifying neural networks,'' \emph{Proc. ACM Program. Lang.}, vol.~3, no.
  POPL, pp. 1--30, 2019.

\bibitem{sun2022causality}
B.~Sun, J.~Sun, L.~H. Pham, and J.~Shi, ``Causality-based neural network
  repair,'' in \emph{Proceedings of the 44th International Conference on
  Software Engineering}, 2022, pp. 338--349.

\bibitem{Sun2018}
Y.~Sun, M.~Wu, W.~Ruan, X.~Huang, M.~Kwiatkowska, and D.~Kroening, ``Concolic
  testing for deep neural networks,'' in \emph{Proceedings of the 33rd ACM/IEEE
  International Conference on Automated Software Engineering}, 2018, p.
  109–119.

\bibitem{supply_chain}
X.~Tan, K.~Gao, M.~Zhou, and L.~Zhang, ``An exploratory study of deep learning
  supply chain,'' in \emph{Proceedings of the IEEE/ACM 44th International
  Conference on Software Engineering}, 2022, pp. 86--98.

\bibitem{tang2021empirical}
Y.~Tang, R.~Khatchadourian, M.~Bagherzadeh, R.~Singh, A.~Stewart, and A.~Raja,
  ``An empirical study of refactorings and technical debt in machine learning
  systems,'' in \emph{Proceedings of the IEEE/ACM 43rd International Conference
  on Software Engineering}, 2021, pp. 238--250.

\bibitem{Tao2020}
G.~Tao, S.~Ma, Y.~Liu, Q.~Xu, and X.~Zhang, ``Trader: Trace divergence analysis
  and embedding regulation for debugging recurrent neural networks,'' in
  \emph{Proceedings of the ACM/IEEE 42nd International Conference on Software
  Engineering}, 2020, p. 986–998.

\bibitem{Tian2018}
Y.~Tian, K.~Pei, S.~Jana, and B.~Ray, ``Deeptest: Automated testing of
  deep-neural-network-driven autonomous cars,'' in \emph{Proceedings of the
  40th International Conference on Software Engineering}, 2018, p. 303–314.

\bibitem{Toledo2021}
F.~Toledo, D.~Shriver, S.~Elbaum, and M.~B. Dwyer, ``Distribution models for
  falsification and verification of dnns,'' in \emph{Proceedings of the 36th
  IEEE/ACM International Conference on Automated Software Engineering}, 2021,
  p. 317–329.

\bibitem{configurable_system}
M.~Velez, P.~Jamshidi, N.~Siegmund, S.~Apel, and C.~K{\"{a}}stner, ``On
  debugging the performance of configurable software systems: Developer needs
  and tailored tool support,'' in \emph{Proceedings of the IEEE/ACM 44th
  International Conference on Software Engineering}, 2022, pp. 1571--1583.

\bibitem{TED}
V.~Vlasov, J.~E.~M. Mosig, and A.~Nichol, ``Dialogue transformers,''
  \emph{CoRR}, vol. abs/1910.00486, 2019.

\bibitem{wen-etal-2015-semantically}
T.-H. Wen, M.~Ga{\v{s}}i{\'c}, N.~Mrk{\v{s}}i{\'c}, P.-H. Su, D.~Vandyke, and
  S.~Young, ``Semantically conditioned {LSTM}-based natural language generation
  for spoken dialogue systems,'' in \emph{Proceedings of the Conference on
  Empirical Methods in Natural Language Processing}, 2015, pp. 1711--1721.

\bibitem{williams2016dialogstate}
J.~Williams, A.~Raux, and M.~Henderson, ``The dialog state tracking challenge
  series: A review,'' \emph{Dialogue \& Discourse}, vol.~7, no.~3, pp. 4--33,
  2016.

\bibitem{fix_that_fails}
R.~Wu, C.~Guo, A.~Y. Hannun, and L.~van~der Maaten, ``Fixes that fail:
  Self-defeating improvements in machine-learning systems,'' in
  \emph{Proceedings of the 35th Conference on Neural Information Processing
  Systems}, 2021, pp. 11\,745--11\,756.

\bibitem{yang2019xlnet}
Z.~Yang, Z.~Dai, Y.~Yang, J.~Carbonell, R.~R. Salakhutdinov, and Q.~V. Le,
  ``Xlnet: Generalized autoregressive pretraining for language understanding,''
  \emph{Advances in neural information processing systems}, vol.~32, 2019.

\bibitem{Yokoyama2019}
H.~Yokoyama, ``Machine learning system architectural pattern for improving
  operational stability,'' in \emph{Proceedings of the IEEE International
  Conference on Software Architecture Companion}, 2019, pp. 267--274.

\bibitem{npm_technical_lag}
A.~Zerouali, E.~Constantinou, T.~Mens, G.~Robles, and J.~Gonz{\'a}lez-Barahona,
  ``An empirical analysis of technical lag in npm package dependencies,'' in
  \emph{Proceedings of the 17th International Conference on Software Reuse},
  2018, pp. 95--110.

\bibitem{ml_testing}
J.~M. Zhang, M.~Harman, L.~Ma, and Y.~Liu, ``Machine learning testing: Survey,
  landscapes and horizons,'' \emph{IEEE Transactions on Software Engineering},
  vol.~48, no.~1, pp. 1--36, 2022.

\bibitem{Zhang2016}
J.~Zhang, X.~Jing, W.~Zhang, H.~Wang, and Y.~Dong, ``Improve the quality of arc
  systems based on the metamorphic testing,'' in \emph{Proceedings of the
  International Symposium on System and Software Reliability}, 2016, pp.
  137--141.

\bibitem{Zhang2020}
P.~Zhang, J.~Wang, J.~Sun, G.~Dong, X.~Wang, X.~Wang, J.~S. Dong, and T.~Dai,
  ``White-box fairness testing through adversarial sampling,'' in
  \emph{Proceedings of the ACM/IEEE 42nd International Conference on Software
  Engineering}, 2020, p. 949–960.

\bibitem{interpretability}
Y.~Zhang, P.~Tiňo, A.~Leonardis, and K.~Tang, ``A survey on neural network
  interpretability,'' \emph{IEEE Transactions on Emerging Topics in
  Computational Intelligence}, vol.~5, no.~5, pp. 726--742, 2021.

\bibitem{zhang2020recent}
Z.~Zhang, R.~Takanobu, Q.~Zhu, M.~Huang, and X.~Zhu, ``Recent advances and
  challenges in task-oriented dialog systems,'' \emph{Science China
  Technological Sciences}, vol.~63, no.~10, pp. 2011--2027, 2020.

\end{thebibliography}
}

\end{document}